\documentclass[times,authoryear]{elsarticle}

\usepackage{jasr}
\usepackage{txfonts}
\usepackage{latexsym}
\usepackage[utf8]{inputenc}
\usepackage[T1]{fontenc}
\usepackage{graphicx}
\usepackage{epstopdf}

\usepackage[switch]{lineno}

\usepackage{url}
\usepackage{xcolor}
\definecolor{newcolor}{rgb}{.8,.349,.1}

\usepackage[citebordercolor=white]{hyperref}
\newcommand{\apj}{ApJ}
\newcommand{\apjl}{ApJL}

\newcommand{\solphys}{SoPh}

\newcommand{\nat}{Nature}
\newcommand{\aap}{A\&A}         
\newcommand{\mnras}{MNRAS}      
\newcommand{\apjs}{ApJS}        
\journal{Advances in Space Research}

\begin{document}

\verso{Hegde \textit{et al.}}
\begin{frontmatter}

\title{Multi-height Diagnosis of MHD Wave Periods and their Propagation in Solar Plages Using IRIS Observations}%

\author{Gayathri \snm{Hegde}}
\author{Pradeep \snm{Kayshap}\corref{cor1}}
\cortext[cor1]{Corresponding author: P. Kayshap}

\affiliation{School of Advanced Sciences and Languages (SASL), VIT Bhopal University, Kothari, Asta, Bhopal, Madhya Pradesh, India - 466114}


\begin{abstract}
This work aims to investigate {\bf magnetohydrodynamic (MHD) waves} in solar plages at five distinct heights, spanning from the photosphere to the upper chromosphere, using spectroscopic observations provided by Interface Region Imaging Spectrograph (IRIS). The dominant period is found to not change within the plages, while, in pores, the dominant period decreases linearly from $\sim$4.20 to $\sim$3.20 minutes from the photosphere to the upper chromosphere. Furthermore, in the solar plages, cross-wavelet analysis reveals that periods from $\sim$ 2.0 to 6.0 minutes propagate till the upper chromosphere from the photosphere. The periods beyond 6 minutes have a zero/constant phase difference (i.e., $\Delta$$\phi$) in the photosphere and in the middle chromosphere. Thus, the 6.0-minute period would be considered the cutoff period at these heights. Next, the propagation speeds of MHD waves above solar plages are estimated in the photosphere and chromosphere. Within the limit of uncertainties, the propagation speed in solar plage is close to the sound speed; hence, the waves are slow magnetoacoustic in nature. Lastly, with the help of $\Delta$$\phi$ analysis, we found that formation heights of Mg~{\sc ii}~k2r and Mg~{\sc ii}~k3 are underestimated in solar plage, while the formation height of Mn~{\sc i} is overestimated. In case of pores, the formation heights of Mg~{\sc ii}k2r, Mg~{\sc ii} k3, and Mn~{\sc i} are overestimated. Interestingly, in the quiet-Sun (QS), the formation heights of Mn~{\sc i} and Fe~{\sc i} are nearly the same, and also the formation height of Mg~{\sc ii}~k3 is similar to the formation height of Mg~{\sc ii}~k2r. In conclusion, some important findings are reported in this work, namely, (1) dominant periods at five different heights between the photosphere and chromosphere, (2) estimation of cutoff periods in the photosphere, middle chromosphere, and the upper chromosphere in the plages, and (3) variations in the formation heights of these spectral lines in pore, plage, and QS.        
\end{abstract}


\end{frontmatter}

\section{Introduction}
\label{S-Intro}
The solar atmosphere comprises various regions such as the active region (AR), sunspot, QS, coronal hole (CH), plages, and many more. The MHD waves are ubiquitous in all these regions of the solar atmosphere; see the review articles by \cite{2015RSPTA.37340261A}, \cite{2015LRSP...12....6K}, and \cite{2023LRSP...20....1J}. The photospheric power spectrum of MHD waves in most of the regions of the solar atmosphere (e.g., QS, umbra, and CH) is dominated by 5 minutes. On the contrary, the power spectrum at the chromosphere of those regions is dominated by 3 minutes (e.g., \citealt{1972SoPh...24...87B, 1979ApJ...231..570L, 1982ApJ...253..367L, 2004ApJ...617..623B, 2006ApJ...640.1153C, 2009ApJ...692.1211C, 2011A&A...534A..65S, 2015LRSP...12....6K, 2017ApJS..229...10J, 2018MNRAS.479.5512K, 2022MNRAS.517..458S, 2024ApJ...966..187S}). However, unlike the other atmospheric regions, the photospheric as well as chromospheric power spectrum of MHD waves in the solar plages and networks are dominated by 5 minutes (e.g., \citealt{1982ApJ...253..367L, 1993ApJ...414..345L, 2003ApJ...595L..63D, 2007ApJ...654.1128D, 2020A&A...634A..63K}).\\

The plages are bright and strong magnetic field regions in the chromosphere, typically found surrounding sunspots/pores. Similar to plages, the networks are also regions of the strong magnetic field, and such regions of a strong magnetic field (i.e., plages and networks) contain magnetic flux tubes (\citealt{2000ApJ...535..489B}). The flux tube can have any orientation with respect to the vertical, i.e., the existence of inclined magnetic flux tubes. The inclined flux tube can increase the cutoff period (or reduce the cutoff frequency; \citealt{2004Natur.430..536D, 2006MNRAS.372..551S}), as a result, the longer periods can reach the transition region (TR)/lower corona in these regions (e.g., \citealt{2005ApJ...624L..61D, 2009ApJ...692.1211C, 2020A&A...634A..63K}). More recently, in a numerical simulation, \cite{2021A&A...652A..43Y} has shown that inside the plage regions the significant fraction of low-frequency waves (i.e., longer periods) leak into the chromosphere due to the inclined flux tubes, and they carry enough heat to the plage chromosphere.\\

A significant amount of wave power within the solar chromosphere is transferred from the photosphere, i.e., chromospheric oscillations are directly related to photospheric oscillations. The study of the wave propagation from one layer to another is the best means to know the wave's contribution in the chromosphere from the photosphere. We mention that a significant amount of work on wave propagation in different regions has been done using the Wavelet/Fourier analysis (e.g., \citealt{1979ApJ...231..570L, 1981SoPh...69..233W, 1982ApJ...253..367L, 1993ApJ...414..345L, 2003ApJ...595L..63D, 2004Natur.430..536D, 2005ApJ...624L..61D, 2007ApJ...654.1128D,2016ApJ...819L..23W, 2017ApJS..229...10J, 2018A&A...617A..39F,2018MNRAS.479.5512K,2020A&A...634A..63K, 2022MNRAS.517..458S, 2024ApJ...966..187S}). However, it should also be noted that wave propagation within the solar plages is the least explored subject, and only a few research works are available on the same (e.g., \citealt{2003ApJ...595L..63D, 2006ASPC..358..465C, 2009ApJ...692.1211C, 2020A&A...634A..63K}). Using the Transition Region and Coronal Explorer (TRACE), it has been demonstrated that longer periods can extend up to the TR from the photosphere in solar plages (\citealt{2003ApJ...595L..63D}). Furthermore, it has shown the propagation of longer periods to the solar chromosphere from the photosphere in the facular regions (e.g., \citealt{2006ASPC..358..465C, 2009ApJ...692.1211C}). More recently, using imaging observations from the IRIS and the Atmospheric Imaging Assembly (AIA), it has been reported that periods from 2 to 9 minutes can reach TR from the high photosphere (e.g., \citealt{2020A&A...634A..63K}).\\

Here, it would be important to note that the cutoff period puts constraints on the wave propagation, as not all waves reach the higher layers. Hence, variations in the cutoff period (frequency) affect the period at different heights of the solar atmosphere. In  QS, \cite{1979ApJ...231..570L} showed that oscillations below the frequency of 4 mHz become evanescent, but the frequencies higher than this would reach higher heights in the solar atmosphere, i.e., the cutoff frequency is 4 mHz in the QS. The same cutoff frequency (i.e., 4 mHz) in the QS is again reported by \cite{2018MNRAS.479.5512K}. Furthermore, \cite{2009ApJ...692.1211C} estimated the cutoff frequencies in various regions, namely sunspots, facular region, and pores. They reported that umbra/pores have a cutoff frequency of 4 mHz, while the facular region has a cut-off frequency of $\sim$2 mHz. In an interesting work, \cite{2016ApJ...819L..23W} estimated cutoff frequencies at multiple heights within the QS. To the best of our knowledge, there is no dedicated observational work focused on the estimation of the cutoff frequencies at multiple heights within plages. Also, as described earlier, only a few works report the dominant periods and their propagation within the different layers of the atmosphere above the solar plages. Hence, more observational works are required to understand various aspects of MHD waves in solar plages, namely, (a) dominant periods at multiple heights, (b) propagation of waves in different atmospheric layers, and (c) cutoff frequencies at different heights.\\

This work investigates some important aspects of MHD waves (e.g., dominant period, propagation, and cutoff frequencies) in the solar plages at five atmospheric heights from the photosphere to chromosphere. In addition to the solar plage, the dominant period is also estimated at five different heights in the pores. The goals of this work are achieved with the help of high-resolution spectroscopic observations of multiple spectral lines. The detailed wavelet analysis is applied to the Doppler velocity time series (DTS, hereafter) to estimate the power spectrum of MHD waves at different heights in the plages/pores. This is the first work to report the variations in the power spectrum at five heights (between the photosphere and the chromosphere) in the solar plage. The paper is organized as follows. Section~\ref{S-obs} describes the observation and data analysis, while the observational results are described in Section~\ref{S-analysis}. The summary and discussions are outlined in the last section. 

\section{Observation and Data Analysis}
\label{S-obs}

IRIS (\citealt{2014SoPh..289.2733D}) provides high-resolution spectroscopic and imaging observations of the solar atmosphere. On June $18^{th}$, 2018, IRIS has observed an AR plage located very close to the disk centre (i.e., x = 66$"$ and y = 72$"$), and the maximum field of view (FOV) of the observed region is 119$"$$\times$119$"$. The observation takes place in sit-and-stare mode from 16:49:50~UT to 21:59:15~UT. The sit-and-stare mode means that the slit is placed at a particular spatial location to capture spectra from that position over a specific period of time. The wave-based studies can be performed using such observations. The IRIS observed AR plage in near-ultraviolet (NUV) and far-ultraviolet (FUV) spectra with an exposure (cadence) time of 15 (16.5) s. The NUV and FUV spectral bands contain several emission and absorption lines that provide vital information about the properties of the plasma in their region of formation. Fe~{\sc i}~2793.22 Å, Mn~{\sc i}~2801.901 Å, Mg~{\sc ii}~k~2796.3 Å, Mg~{\sc ii}~h~2803.5 Å, C~{\sc ii}~1336 Å, Si~{\sc iv}~1403 are just few spectral lines that form from the photosphere to the TR (\citealt{2014SoPh..289.2733D}). \\

In addition to spectroscopic observations, IRIS also provides imaging observations (i.e., slit-jaw images (SJIs)) in various wavebands, namely SJI~2832~{\AA}, SJI~2796~{\AA}, SJI~1400~{\AA}, SJI~1330~{\AA} (\citealt{2014SoPh..289.2733D}). In this observation, the full FOV of the SJI is 119$"$$\times$119$"$, and the cadence of SJI is 66 s. Further, we have also utilized line-of-sight (LOS) magnetic field observations from the Helioseismic and Magnetic Imager (HMI; \citealt{2012SoPh..275..207S}) onboard Solar Dynamic Observatory (SDO; \citealt{2012SoPh..275....3P}). We aligned the absolute LOS magnetic field with the chromospheric intensity (i.e., IRIS/SJI Mg~{\sc ii} k 2796~{\AA}). As the work aims to understand the dominant periods at multiple heights and their propagation within the different layers, we have considered three different lines for this study, namely Fe~{\sc i}~2793.22~{\AA}, Mn~{\sc i}~2801.901~{\AA}, and Mg~{\sc ii}~k~2796~{\AA}. Note that Mg~{\sc ii}~k line is an optically thick line, and usually it has two wings (i.e., k2v and k2r) and a central dip region (i.e., k3) that forms at different heights within the solar atmosphere (\citealt{1981ApJS...45..635V, 2013ApJ...772...90L}). Hence, with Mg~{\sc ii} k line, we have the advantage of diagnosing the waves at the formation heights of Mg~{\sc ii} k$_{2r}$, Mg~{\sc ii} k$_{2v}$ and Mg~{\sc ii} k$_{3}$. Lastly, the formation heights of all these lines and corresponding uncertainties are mentioned in table~\ref{tab:tab1}. As per \cite{1981ApJS...45..635V}, Mg~{\sc ii}k2r and Mg~{\sc ii}k2v forms nearly at the same height. However, on the other hand, \cite{2013ApJ...772...90L} reported that Mg~{\sc ii} k2v forms slightly higher than Mg~{\sc ii}k2r.

\begin{table*}
\centering
\caption{ Formation heights of the spectral lines
}
\label{tab:tab1}
\begin{tabular}{|c|ccc|c|c|}  

\hline
Sl.No &  &  Spectral Line &  &Formation Height &  Uncertainty\\
      &   &   &  &(Mm) &  (Mm)\\
  \hline
1 &  &  Fe~{\sc i} &  &0.50&   0.04  \\ 
\hline
2 &  & Mn~{\sc i} &  &0.83& 0.10 \\
\hline
3 &  &  Mg~{\sc ii}~$k_{2r}$ &  & 1.40 &  0.30  \\
\hline
4 &  & Mg~{\sc ii}~$k_{2v}$ &  & 1.40 &  0.30 \\
\hline
5 &  & Mg~{\sc ii}~$k_{3}$ &  & 1.90 &  0.10\\
\hline
\end{tabular}
\end{table*}

Next, we have deduced the Doppler velocity of the two absorption lines (e.g., Fe~{\sc i}~2793.22~{\AA} and Mn~{\sc i}~2801.901~{\AA}) by applying the Gaussian fit over the observed profiles. This fit allows for the estimation of centroids and Gaussian width at each spatial location, and further, the centroids are converted into the Doppler velocities with the help of rest wavelengths. For the Mg~{\sc ii}~k line, it must be noted that the single Gaussian fit is not appropriate, as it is an optically thick line (\citealt{2013ApJ...772...89L}). We have used the interactive data language (IDL) routine (i.e., iris$\_$get$\_$mg$\_$features$\_$lev2.pro) to obtain the Doppler velocities of Mg~{\sc ii}~k2r, Mg~{\sc ii}~k2v, and Mg~{\sc ii}~k3. The details of this method are described by \cite{2013ApJ...778..143P}.
\section{Observational Results}
\label{S-analysis}

\subsection{Region of Interest (ROI)} \label{sect:roi}
\begin{figure}
\centering
\includegraphics[trim=0.0cm 0.0cm 0.0cm 2.0cm,scale=1.0]{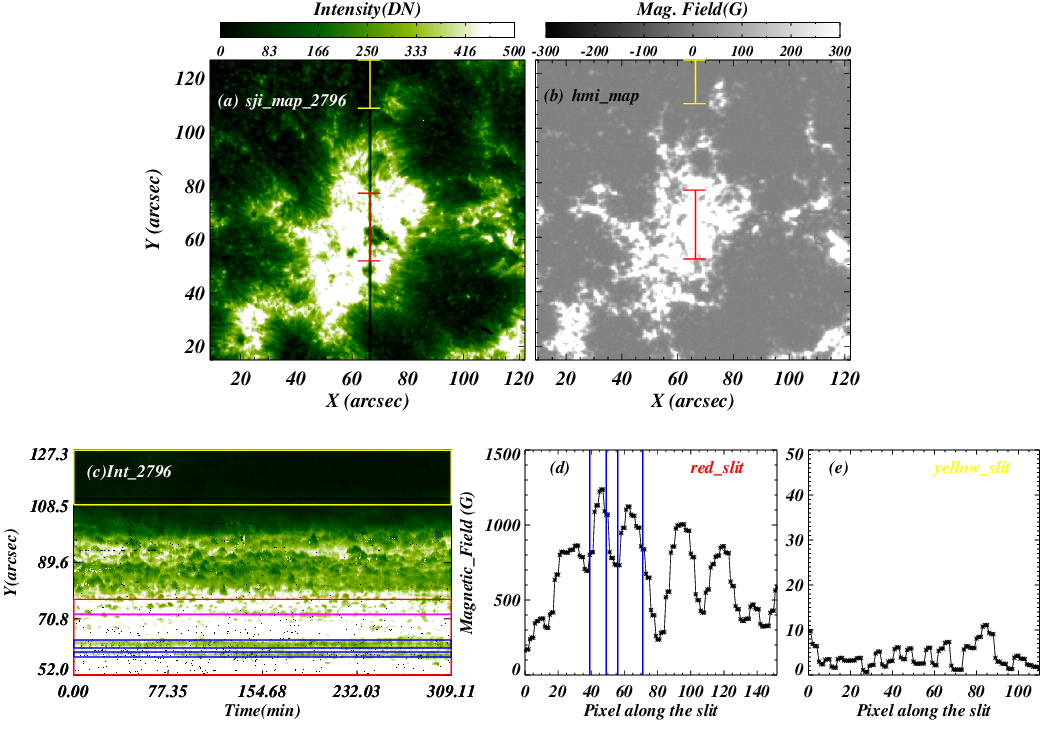}
\caption{The panel (a) displays the IRIS/SJI Mg~{\sc ii}~k~2796~{\AA}(chromospheric emission) image of the solar plage. The LOS magnetogram of the observed region is shown in panel (b). The black vertical line in panel (a) represents the IRIS slit position. The red line drawn over the slit in panels (a) and (b) covers the significant region of plage. Along with this, it is also covering the pore region. The yellow slits drawn in panels (a) and (b) are the QS region. Panel (c) is the spectral intensity image of Mg~{\sc ii}~k~2796~{\AA} along all the 'y' locations from plage to QS. The QS area is represented in the yellow box, the pore regions are represented in two blue boxes, and the plage area for study is represented in the red box. The magnetic field strength along the red and yellow slits is shown in panels (d) and (e). The pore (area enclosed by vertical blue lines in panel (d)) has a higher magnetic field than the plage area. The QS region shows very low magnetic strength (panel (e)).}
\label{fig:ref_fig}
\end{figure}

Figure~\ref{fig:ref_fig} displays the IRIS/SJI Mg~{\sc ii}~k~2796~{\AA} image (chromospheric emission, panel (a)) and the LOS magnetogram (panel (b)) of the observed region. A big, bright intensity patch is visible in the solar chromosphere, and the LOS magnetogram shows that this bright intensity patch belongs to a unipolar strong magnetic field patch. The enhanced emission in the chromosphere and strong unipolar magnetic field patch at the solar photosphere are signatures of solar plages. Hence, we mention that this particular region is an AR solar plage, which is the region of interest (ROI) of this study. Some small dark regions exist within this bright intensity patch, and most probably these dark regions are pores.\\

The black vertical line in panel (a) represents the IRIS slit position, and IRIS captures the spectra from the region underneath the slit. Firstly, we have selected a portion of the slit that lies mostly in the plage region, i.e., the region covered by the red slit in panel (a). Further, the region north of the plage, covered by the yellow slits, is the QS. The region covered by these red and yellow slits is also displayed on the HMI magnetogram (panel b), and these regions are our ROI. Most of the area under the red slit is the plage, but a small fraction of the red silt crosses the dark regions (i.e., pores). Now, before moving further, we must segregate the solar plage and pore regions.\\

Mg~{\sc ii}~k3 spectral intensity (derived from the chromosphere) is used to segregate pore and plage. The Mg~{\sc ii} k3 spectral intensity map is shown in panel (c). The lower and upper coordinates of the red slit are 52.0$"$ and 76.4$"$. Using this information, we have overplotted a red rectangular box in panel (c), which encloses only the plage region. Here, we have intensity time-series at each Y position of the red slit (within the plage). Further, the intensity time-series is averaged over all Y positions, and in this way, we have a total of 152 averaged intensity points between Y = 52.0" and Y = 76.40", i.e., 152 locations along the red slit only. As explained in the next paragraph, the least 22\% intensity can be considered as intensity threshold for pore selection. Using this intensity threshold, we identified all locations, which are enclosed by two blue rectangular boxes in panel (c). Note that the regions enclosed by the two blue boxes have lower intensity in comparison to the surrounding plage region, which justifies that the chosen regions are the pores inside the plage. To strengthen this, we have plotted the LOS magnetic field along the red slit in panel (d). Further, the pore locations are enclosed by vertical blue lines in panel (d), and it is found that the strength of the magnetic field inside the blue boxes (i.e., pore) is higher than the strength of the magnetic field outside of the blue boxes (i.e., plage). Hence, the lower intensity and higher magnetic field in the locations outlined by the blue boxes justify the fact that these locations are pore regions. Further, we have also shown the magnetic field along the yellow slit (i.e., QS) in panel (e), and the magnetic field is very weak, which justifies that the selected region is QS.\\

The dominant periods are the same in the pore and plage in the photosphere, but are different in the chromosphere (\citealt{2009ApJ...692.1211C}). Also, we know that the spectral intensity would be lower in the pores than in the plage at the photosphere and chromosphere. Hence, the intensity and dominant periods in the pores differ from those in plage at the chromospheric level. Therefore, to isolate pores and plages, we used spectral intensity and dominant periods from chromospheric Mg~{\sc ii} k3~2796.35~{\AA}. The panel (c) of Figure~\ref{fig:ref_fig} is the Mg~{\sc ii} k3 spectral intensity map, and the red rectangular box encloses the solar plage. Within the red box, we estimate the mean intensity from each intensity time series. Next, we have chosen all the locations, out of the 152 locations, that have least 6\% mean Mg~{\sc ii} k3 intensity. Similarly, we have obtained sets of locations corresponding to the least 12 \%, 18\%, 22\%, and 25 \% mean intensities. Further, with the help of wavelet analysis, the dominant periods are estimated for each set of locations.
\begin{figure}
\centering
\includegraphics[trim=0.0cm 1.5cm 0.0cm 1.5cm,scale=1.0]{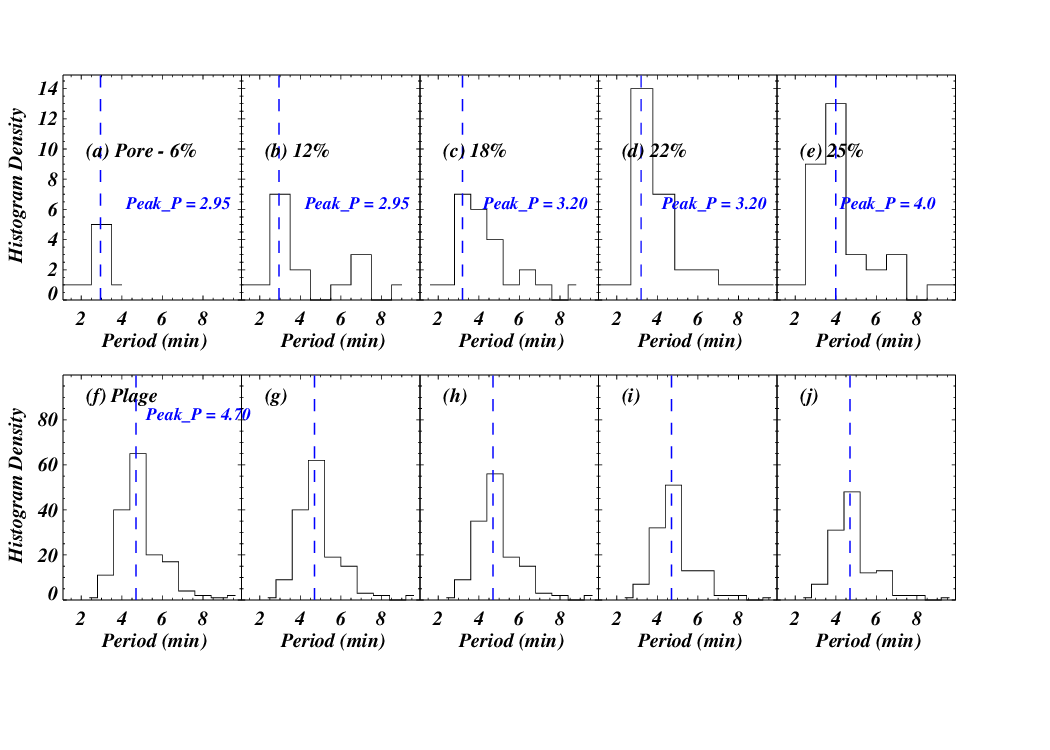}
\caption{The dominant peak periods at Mg~{\sc ii}~k3 height, with different intensity thresholds for classification of plage and pore locations, are shown here. The panels (a) to (e) represent the dominant period for the pore locations. The dominant periods for up to a threshold of 22\%  are found to be $\approx3min$. For an increased intensity threshold of 25\% the period is $\approx4min$. The histogram of dominant periods corresponding to the highest 94\%, 88\%, 82\%, 78\%, and 75\% intensities is displayed in panels (e) to (j), and all the histograms show that the dominant period is the same; $\approx4.7min$.}
\label{fig:pore_plage_select}
\end{figure}
The histograms of the dominant periods collected from the set of locations corresponding to least 6\%, 12\%, 18\%, 22\%, and 25\% Mg~{\sc ii}~k3 intensities are displayed in panels (a), (b), (c), (d), and (e) of Figure~\ref{fig:pore_plage_select}, respectively.\\

The peak dominant periods are 2.95 (panel a), 2.95 (panel b), 3.20 (panel c), 3.20 (panel d), and 4.0 minutes (panel d) for the locations corresponding to least 6\%, 12\%, 18\%, 22\%, and 25\%, respectively. It means that the peak dominant period increases as we increase the intensity threshold from least 6\% to 25\%. The peak period till least 22\% intensity is 3.20, while if we increase it further by just 3\% (which adds only 4 new locations), then the peak period is shifted to 4.0 minutes. We know that for the pore/umbra, the dominant period lies in the range of 3-4 min in the chromosphere (e.g., \citealt{2009ApJ...692.1211C, 2021ApJ...906..121K}). More importantly, these newly added 4 locations have periods longer than 5 minutes, as you see the histogram in panel (e) has a higher number of longer periods than in panel (d). Keeping all these in view, we have considered that the locations corresponding to the least 22\% intensity would be the pore locations. Furthermore, the histograms of dominant periods from the sets of locations corresponding to the highest 94\%, 88\%, 82\%, 78\%, and 75\%  intensities are displayed from panels (f) to (j), respectively. The peak dominant period is nearly the same (i.e., 4.70 minutes) in all the panels because the majority of locations in any set belong to the plage. Consequently, we mention that the locations corresponding to the least 22\% intensities are considered as the pore region, while the rest of the locations (i.e., the remaining 78\%) are considered as the plage region.\\

To strengthen our logic about pore and plage selection, we further estimated the dominant period from QS, i.e., the region enclosed by the yellow box in panel (c) of Figure~\ref{fig:ref_fig}.
\begin{figure}
\centering
\includegraphics[trim=0.0cm 0.5cm 0.0cm 0.5cm,scale=0.9]{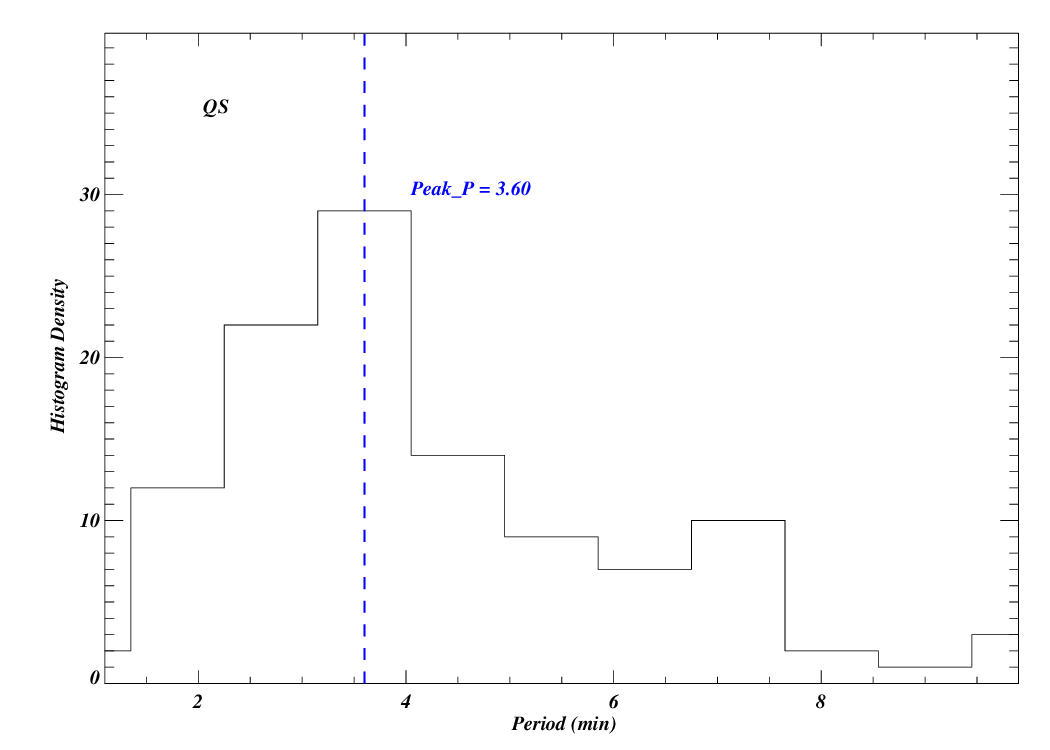}
\caption{{The figure shows the histogram from QS at the chromospheric height}. The dominant period is 3.60 mins.}
\label{fig:qs_k3}
\end{figure}
The wave spectrum from the photosphere of QS and plage dominates by 5 minutes, while the wave spectrum of the chromosphere above QS  and plage is dominated by 3-min  and 5-min (\citealt{2009ApJ...692.1211C, 2018MNRAS.479.5512K, 2020A&A...634A..63K}). The dominant periods in the chromosphere above QS and plage are different. Figure~\ref{fig:pore_plage_select} (panels (f) to (j)) shows that the dominant period in plage at the chromosphere is nearly 5 minutes. While Figure~\ref{fig:qs_k3} shows that the dominant period is around 3.60 minutes. This analysis shows that the region enclosed by the yellow slits (i.e., QS) is different from the plage. Here, it would also be important to note that the dominant period in the chromosphere above the pore/umbra is the same as QS, i.e., 3 minutes (e.g., \citealt{2009ApJ...692.1211C, 2021ApJ...906..121K}). And, again, our analysis shows that the dominant periods in the chromosphere above the pore (Figure~\ref{fig:pore_plage_select}(d)) and QS (Figure~\ref{fig:qs_k3}) are nearly the same. Lastly, we mention that we have successfully segregated the pore and plage regions.

Along the red slit (shown in panels (a) and (b); Figure~\ref{fig:ref_fig}), we have a total of 152 spatial locations (or pixels). Among these, only 30 locations are in the pore region (i.e., the locations corresponding to the lowest 22\% intensity), and the rest of the locations (i.e., 122 locations) are in the plage region. {\bf It is important to note that 22\% of the total 152 locations corresponds to 33. But we found that three locations have comparatively less intensity within the solar plage, as indicated by the magenta lines in panel (c) of Figure~\ref{fig:ref_fig}. To verify whether these locations are indeed the pore, we have checked the dominant periods at these three locations, and they are all around 5 minutes. Therefore, we have considered these three locations as plage locations, not in the pore.} Hence, we have a total of 122 (30) DTS for the plage (pore) at each spectral height. For each DTS, we have 1128 time points (the corresponding time is 1128$\times$16.5 = 18612s $\sim$ 310 min) in sit-and-stare mode observations. Upon analyzing the spectra, we found that the spectra were affected by cosmic hits after 157 min (i.e., after time corresponding to 570 steps) and continued for the next 10 minutes. Later, one more patch of poor-quality spectra exists from time 245 to 259 min. The severity of bad spectra for this time range is discussed in Appendix~A. Therefore, we have considered the initial 570 time points (i.e., $\sim$ 157 minutes) for further analysis.

\subsection{Power at the Solar Photosphere and Chromosphere}
\begin{figure}
\centering
\includegraphics[trim=0.0cm 0.5cm 0.0cm 0.0cm,scale=1.0]{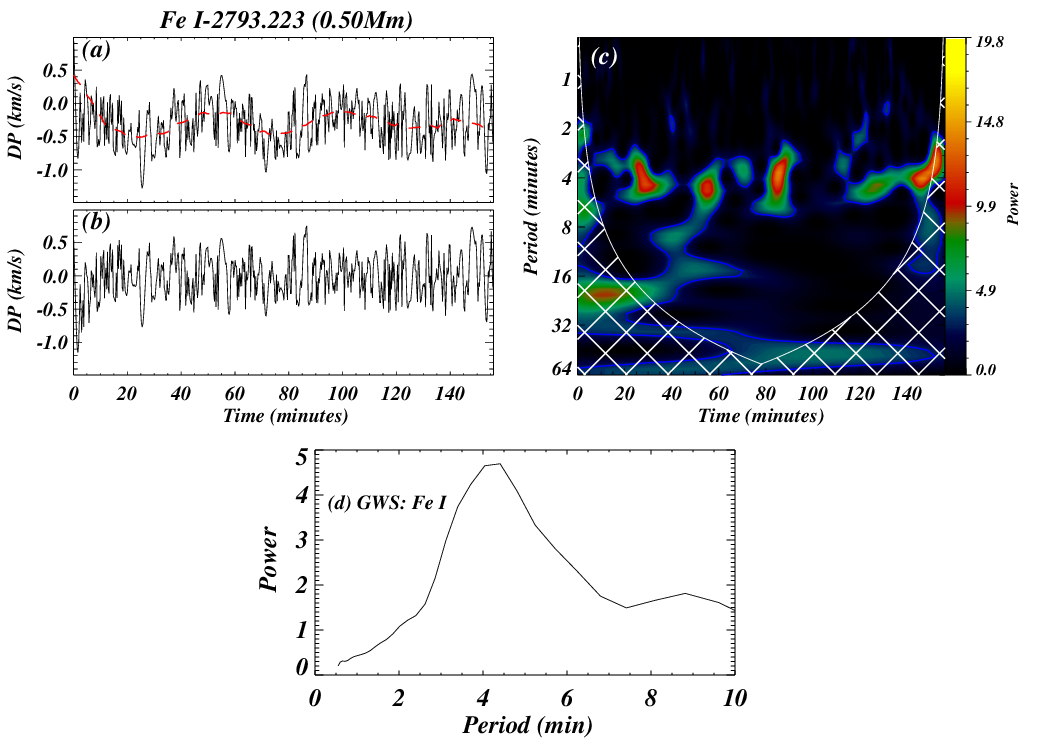}
\caption{The panel (a) represents the DTS of the photospheric Fe~{\sc i} line from a particular location in the solar pore. The smoothed DTS is overplotted by the red curve on the original DTS (black curve). Panel (b) depicts the detrended DTS, i.e., original DTS (black curve) - smoothed DTS (red curve). Panel (c) shows the wavelet power map for detrended DTS. The blue-dashed contours outline wavelet power with 95\% confidence level. The cross-hatched white area represents the COI. The panel (d) represents the global wavelet spectra (GWS), which peaks around 5 minutes.} 
\label{fig:wave_pore_fe}
\end{figure}

Wavelet analysis is an important tool to study waves/oscillations in the time-series (e.g., \citealt{1998BAMS...79...61T}), and it is widely used in various works to study waves in the solar atmosphere(e.g.,\citealt{2004ApJ...617..623B, 2007A&A...473..943J, 2012A&A...539L...4S, 2015SoPh..290..363K, 2017ApJS..229...10J, 2020A&A...634A..63K, 2024ApJ...966..187S}). Figure~\ref{fig:wave_pore_fe} (a) shows the DTS (deduced from Fe~{\sc i}) from a particular location of the pore. First, we obtain the smoothed DTS using a window of 85 points (i.e., time = 85$\times$16.5 = 1402.5 s/60 = 23.38 mins), and this smoothed DTS is overplotted using a red curve in Figure~\ref{fig:wave_pore_fe}(a). Next, the smoothed DTS (red curve) is subtracted from the original DTS, i.e., black curve - red curve. The detrended DTS is displayed in panel (b). The uncertainties (i.e., 1-sigma error) in the Doppler velocities of the photospheric lines are less than 0.03 km/s, whereas, on average, the Doppler velocities of photospheric lines vary from -1.2 to 1.2 km/s (panel (a)), i.e., the total change is around 2.5 km/s, which is much larger than the uncertainties. We have performed the wavelet analysis of detrended DTS shown in Figure~\ref{fig:wave_pore_fe}(a), and the corresponding wavelet power map is shown in panel (c) of Figure \ref{fig:wave_pore_fe}. 
The white cross-hatched area represents the cone of influence (COI), and the blue contours outline the power within the 95\% confidence level. From the wavelet power map, it is clear that most of the significant power is concentrated within the p-mode frequency range (i.e., 3 to 5 minutes). The black curve in panel (d) represents the global wavelet power spectrum (GWS) of the power map shown in panel (c). In this GWS, the power is found to be peaking at roughly five minutes. In the same way, all wavelet power maps and GWS for the 122 spatial locations of plages and the 30 locations of pores are obtained at this spectral height, i.e., the formation height of Fe~{\sc i}. In most of the locations of the pore, power is peaking around three minutes, and some locations are showing multiple peaks in GWS. We estimate the period corresponding to maximum power within the period range from 1 to 10 minutes, and this period corresponding to maximum power is known as the dominant period. The same exercise is performed at all spectral line heights, such as the formation heights of Mn~{\sc i}, Mg~{\sc ii}~k2v, Mg~{\sc ii}~k2r, and Mg~{\sc ii}~k3. This way, we collect all the dominant periods from each spectral height for the pore and plage.\\

\begin{figure}
\centering
\includegraphics[trim=0.0cm 0.5cm 0.0cm 0.0cm,scale=1.0]{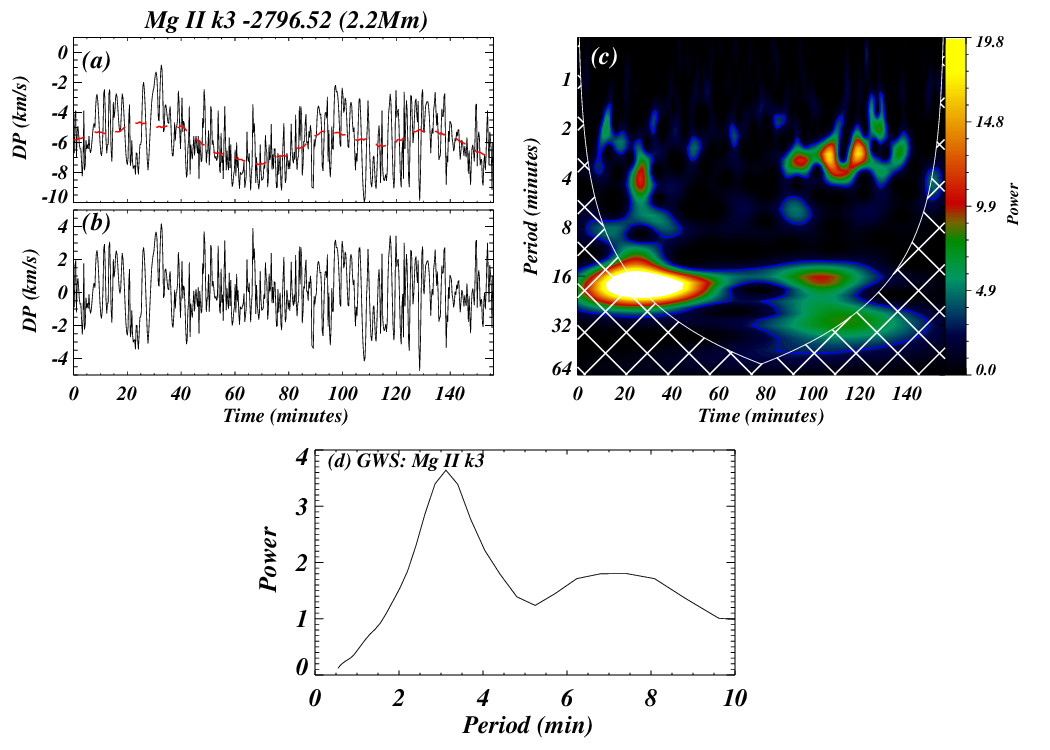}
\caption{Same as Figure~\ref{fig:wave_pore_fe} but for Mg~{\sc ii} k3.}
\label{fig:wave_pore_k3}
\end{figure}

\begin{figure}
\centering
\includegraphics[trim=0.0cm 0.5cm 0.0cm 1.5cm,scale=1.0]{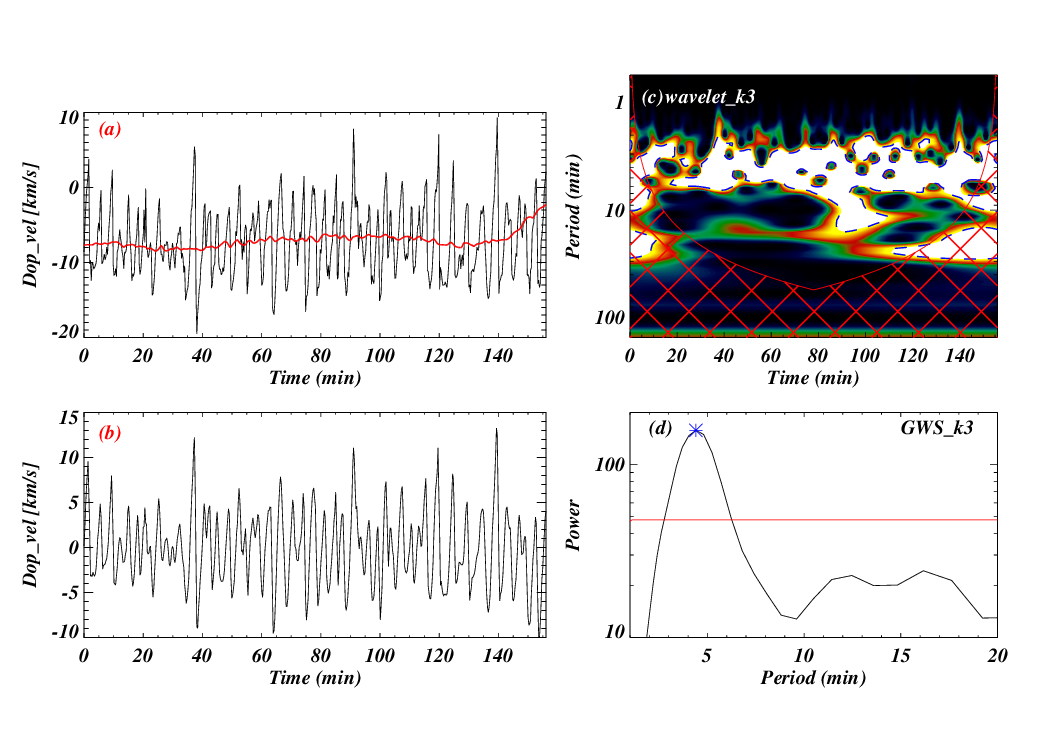}
\caption{Same as Figure~\ref{fig:wave_pore_k3} but for a plage location.}
\label{fig:wave_plage_k3}
\end{figure}

The wavelet power map of Mg~{\sc ii}~k3 (chromosphere) from the same location of the pore as that considered for the photosphere is shown in Figure~\ref{fig:wave_pore_k3}. Panel (a) shows the DTS of the pore location deduced from Mg~{\sc ii}~k3, over which the smoothed (time = 85$\times$16.5 = 1402.5 s/60 = 23.38 mins) red curve is plotted. In panel (b) of Figure~\ref{fig:wave_pore_k3}, the detrended curve is displayed, and it shows that the Doppler velocity varies between -4 and 4km/s. The uncertainties in the Mg II k3 Doppler velocities are less than 1 km/s, i.e., the uncertainties in the DTS in panels (a) and (b) are much smaller.
Next, wavelet analysis is applied to detrended DTS (panel (b)), and the wavelet power map is displayed in panel (c). The white cross-hatched area is the COI, and the blue contours enclose the power within the 95\% confidence level. The significant power is present mainly in periods of $\approx$2 to $\approx$4 min. Also, power is concentrated around 16 minutes, see the bright patch from 0 to 60 minutes around the period of 16 minutes. However, most of the power of this patch lies within the COI. Further, in panel (d), the GWS at this height is shown, and it is evident that the period is roughly peaking at 3 minutes. The location belongs to pore, therefore, the dominant period is around 5 mins at the photosphere (Figure~\ref{fig:wave_pore_fe}(d)) and 3 mins at the chromosphere (Figure~\ref{fig:wave_pore_k3}(d)).
\begin{figure}
\centering
\includegraphics[trim=0.0cm 0.5cm 0.0cm 1.0cm,scale=1.0]{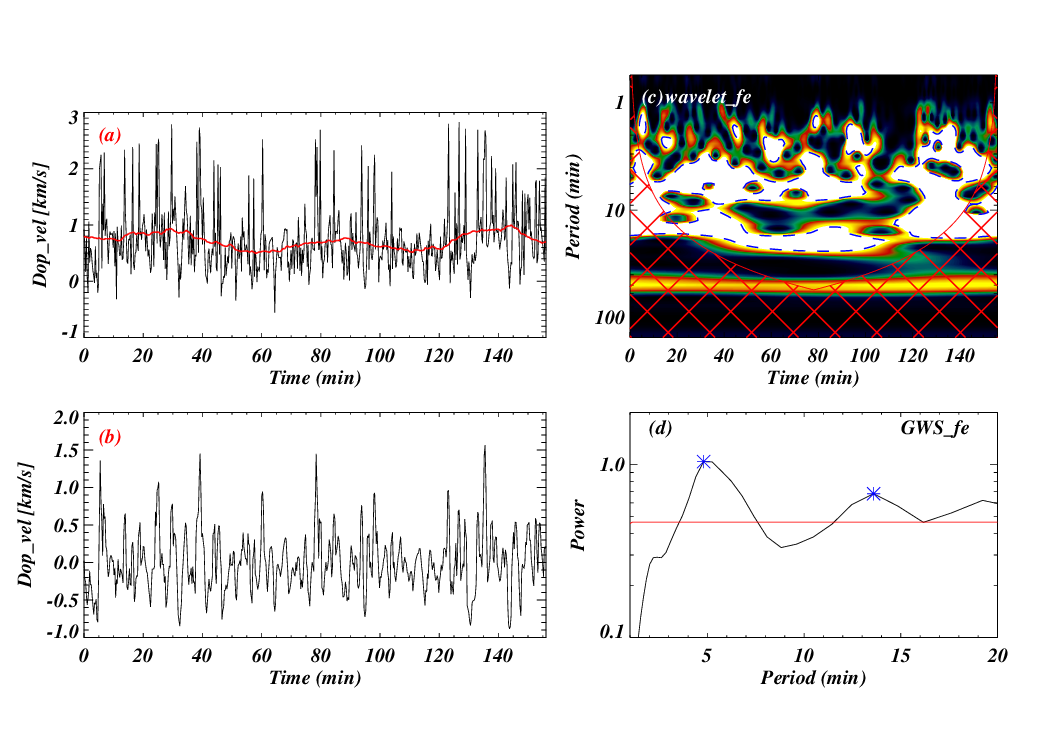}
\caption{Same as Figure~\ref{fig:wave_pore_fe} but for a plage location.}
\label{fig:wave_plage_fe}
\end{figure}
Similar to Figures~\ref{fig:wave_pore_fe} and~\ref{fig:wave_pore_k3}, the wavelet analysis from one particular plage location is displayed for the Fe~{\sc i} (Figure~\ref{fig:wave_plage_fe}) and Mg~{\sc ii} k3 (Figure~\ref{fig:wave_plage_k3}). The majority of the wave power in plage at the photosphere is concentrated around 5 minutes (panels (c) and (d); Figure~\ref{fig:wave_plage_fe}). Additionally, some significant power is also concentrated around 13 minutes, see the elongated big patch above the period of 10 minutes in panel (c) of Figure~\ref{fig:wave_plage_fe}.  Further, the wave power in the chromosphere above solar plage is concentrated around 5 minutes (panels (c) and (d); Figure~\ref{fig:wave_plage_k3}).

\subsection{Statistical Analysis of Dominant Wave Periods at Multiple Heights}
\begin{figure}  
\centering
\includegraphics[trim=0.0cm 2.0cm 0.0cm 1.2cm,scale=0.90]{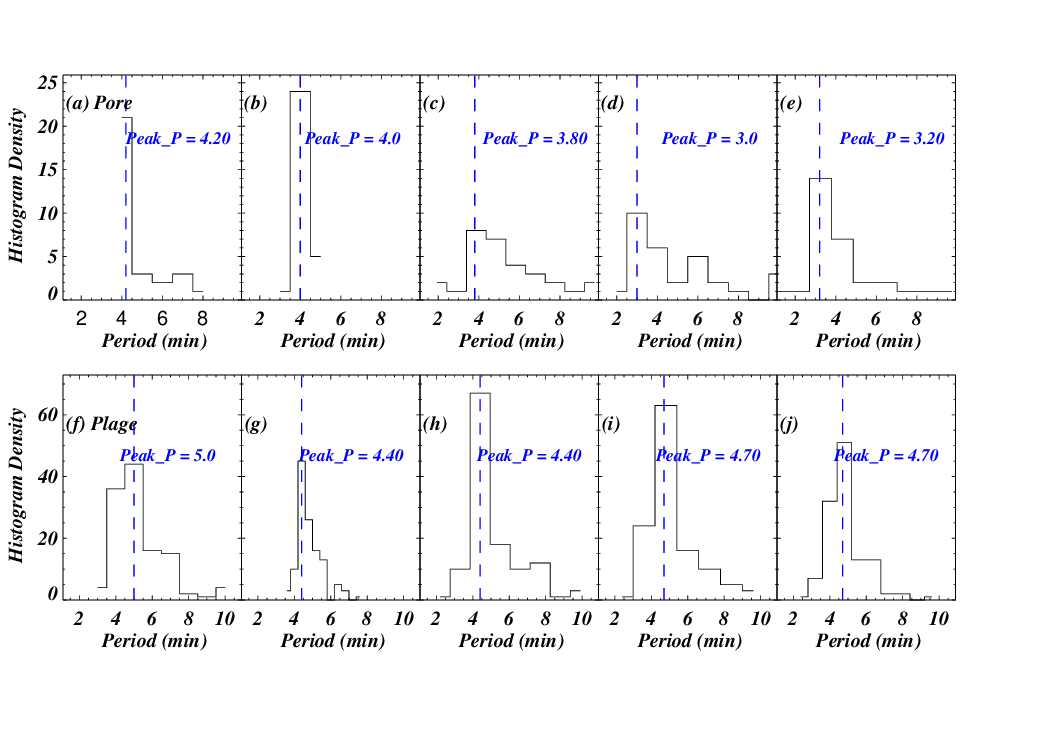}
\caption{The histograms of the dominant periods deduced from all the plage locations from Fe~{\sc i} (panel (a)), Mn~{\sc i} (panel (b)), Mg~{\sc ii} k2r (panel (c)), Mg~{\sc ii} k2v (panel (d)), and Mg~{\sc ii} k3 (panel (e)). The vertical red dashed line in all panels is at the peak dominant period. Further, the same is displayed for the pore regions (i.e., panels (f) to (j)).}
\label{fig:domin_hist}
\end{figure}

\begin{figure}
\centering
\includegraphics[trim=0.0cm 0.0cm 0.0cm 1.2cm,scale=0.90]{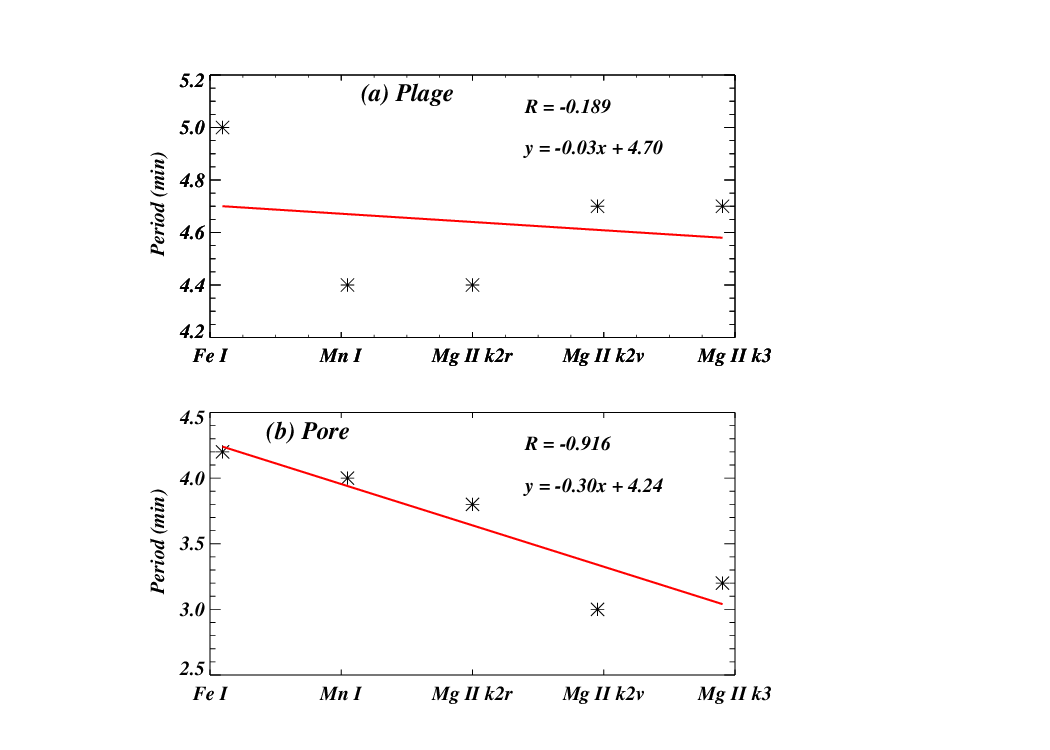}
\caption{The correlation between the peak dominant periods and the spectral line (heights) is represented here for plage (panel (a)) and pore (panel (b)). The dominant periods of plage do not show significant correlation with height, i.e., no change with height. Whereas the same for pore is showing an almost perfect negative correlation (i.e., Pearson coefficient = 0.916), i.e., the dominant periods are decreasing linearly in the pore, unlike the plage.}
\label{fig:domin_correl}
\end{figure}

We collected all the dominant periods from GWS (as displayed in panels (d) of Figure~\ref{fig:wave_pore_fe}, Figure~\ref{fig:wave_pore_k3}, Figure~\ref{fig:wave_plage_fe}, and Figure~\ref{fig:wave_plage_k3}) from all plage (i.e., 122) and pore locations (i.e., 30) at all the spectral lines (heights). After the collection of all dominant periods, we have produced the histogram of all dominant periods at all heights for both pore and plage locations, as shown in Figure~\ref{fig:domin_hist}. 
The histograms in the top row of Figure~\ref{fig:domin_hist} are dedicated to the pore. In panels (a) to (d), the dominant periods are found to peak at 4.2, 4.0, 3.80, 3.0 and 3.20 minutes for Fe~{\sc i}, Mn~{\sc i},  Mg~{\sc ii}~k2r, Mg~{\sc ii}~k2v, and Mg~{\sc ii}~k3, respectively. This indicates that, in the pore, the periods are decreasing {\bf from 4.20 to 3.20 minutes}. The similar period histograms for the plage regions are displayed in the bottom row of Figure~\ref{fig:domin_hist}. The periods in the plage at the formation height of Fe~{\sc i}, Mn~{\sc i},  Mg~{\sc ii}~k2r, Mg~{\sc ii}~k2v, and Mg~{\sc ii}~k3 are 5.0, 4.40, 4.40, 4.70, and 4.70 minutes, respectively. This implies that for plages, the dominant periods do not vary much from the photosphere to the upper chromosphere. The peak periods at all spectral heights for plage and pore are mentioned in Table~\ref{tab:tab2}.
\begin{table*}
\centering
\caption{Periods of Plages and Pores  at different spectra heights
}
\label{tab:tab2}
\begin{tabular}{|c|ccc|c|c|}  

\hline
Sl.No &  &  Spectral Height &  & Peak Period: Plage &  Peak Period: Pore\\
      &   &   &  & (Min) &  (Min)\\
  \hline
1 &  &  Fe~{\sc i} &  & 5.0 &   4.20  \\ 

2 &  & Mn~{\sc i} &  & 4.40 & 4.0 \\

3 &  &  Mg~{\sc ii}~$k_{2r}$ &  & 4.40 &  3.80  \\

4 &  & Mg~{\sc ii}~$k_{2v}$ &  & 4.70 &  3.0 \\

5 &  & Mg~{\sc ii}~$k_{3}$ &  & 4.70 &  3.20 \\

\hline
\end{tabular}
\end{table*}
Furthermore, the peak periods are plotted against the spectral lines (i.e., formation heights of these lines) to understand the correlation between the dominant periods and heights of the solar atmosphere (Figure~\ref{fig:domin_correl}). Panel (a) of this figure shows the variation of dominant periods with height for the plage. It can be seen from the figure that there is no significant variation in the period from the photosphere to the upper chromosphere, as previously explained. The Pearson coefficient for this correlation is just -0.189, i.e., no correlation between dominant period and formation height. 
However, for the pore region, we observe a linear (almost) decrease in the period from $\sim$4.2 to $\sim$3 minutes. The Pearson coefficient (i.e., - 0.916) reveals the strong negative correlation, i.e., the dominant period decreases as height increases. The present results are consistent with the previous findings (e.g.,   \citealt{2006ApJ...640.1153C, 2009ApJ...692.1211C, 2004Natur.430..536D}). 

\subsection{Wave Propagation in the Solar Atmosphere}
\begin{figure}
\centering
\includegraphics[trim=0.0cm 1.2cm 0.0cm 1.0cm,scale=0.3]{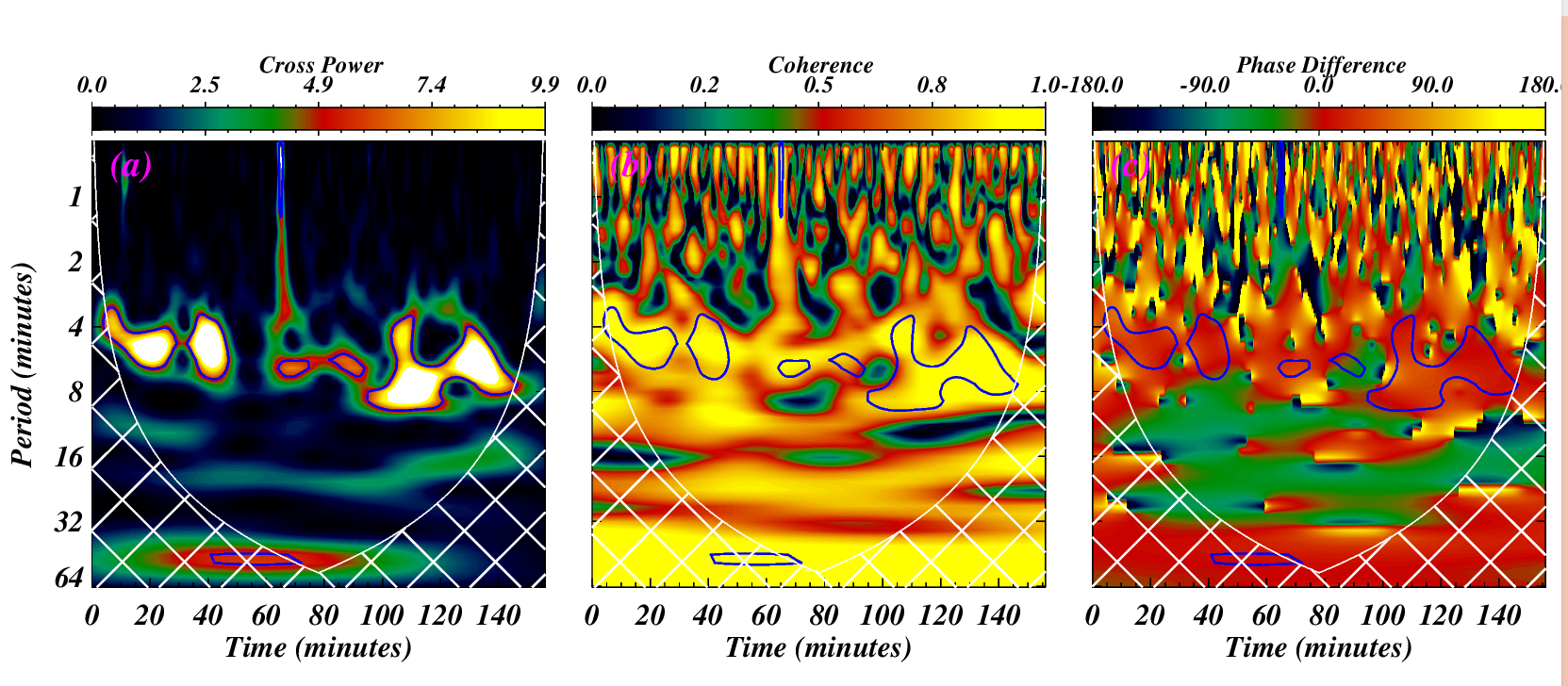}
\includegraphics[trim=0.0cm 1.2cm 0.0cm 0.0cm,scale=0.3]{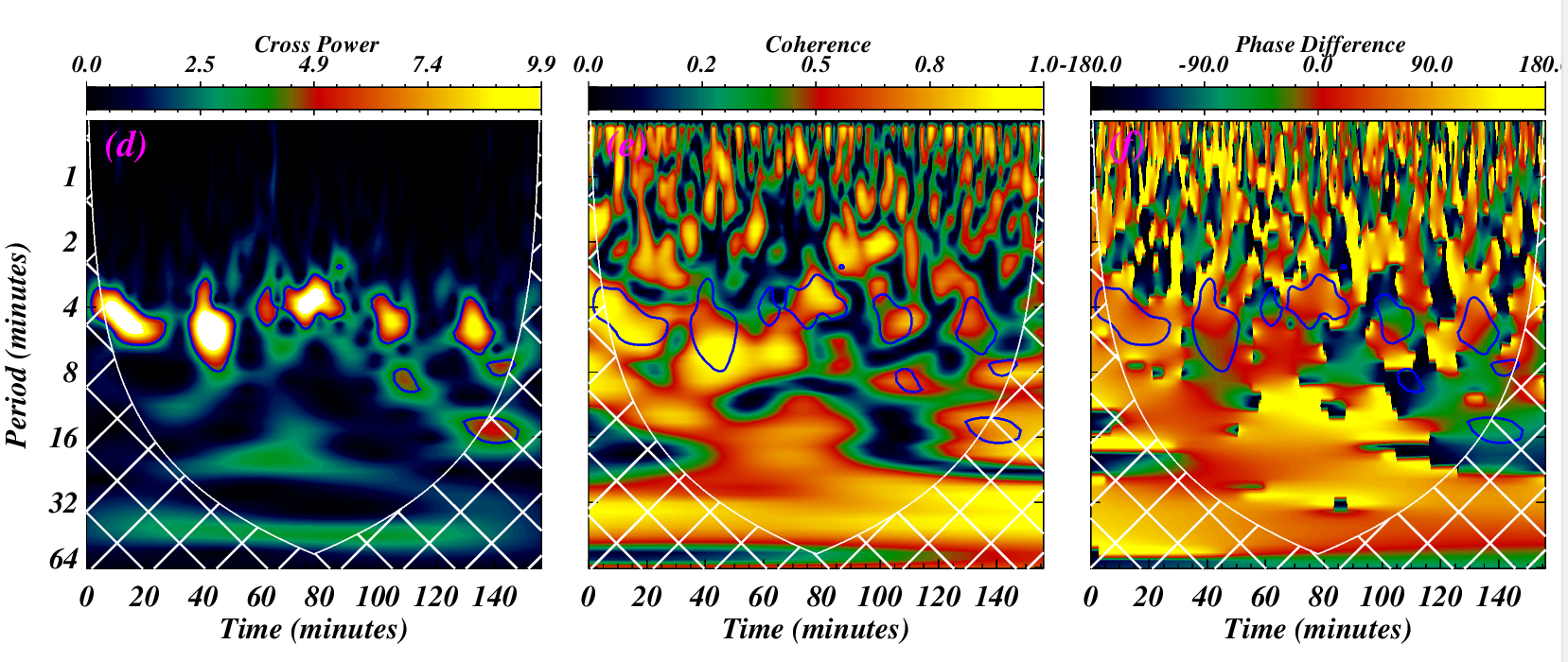}
\includegraphics[trim=0.0cm 1.0cm 0.0cm 0.0cm,scale=0.3]{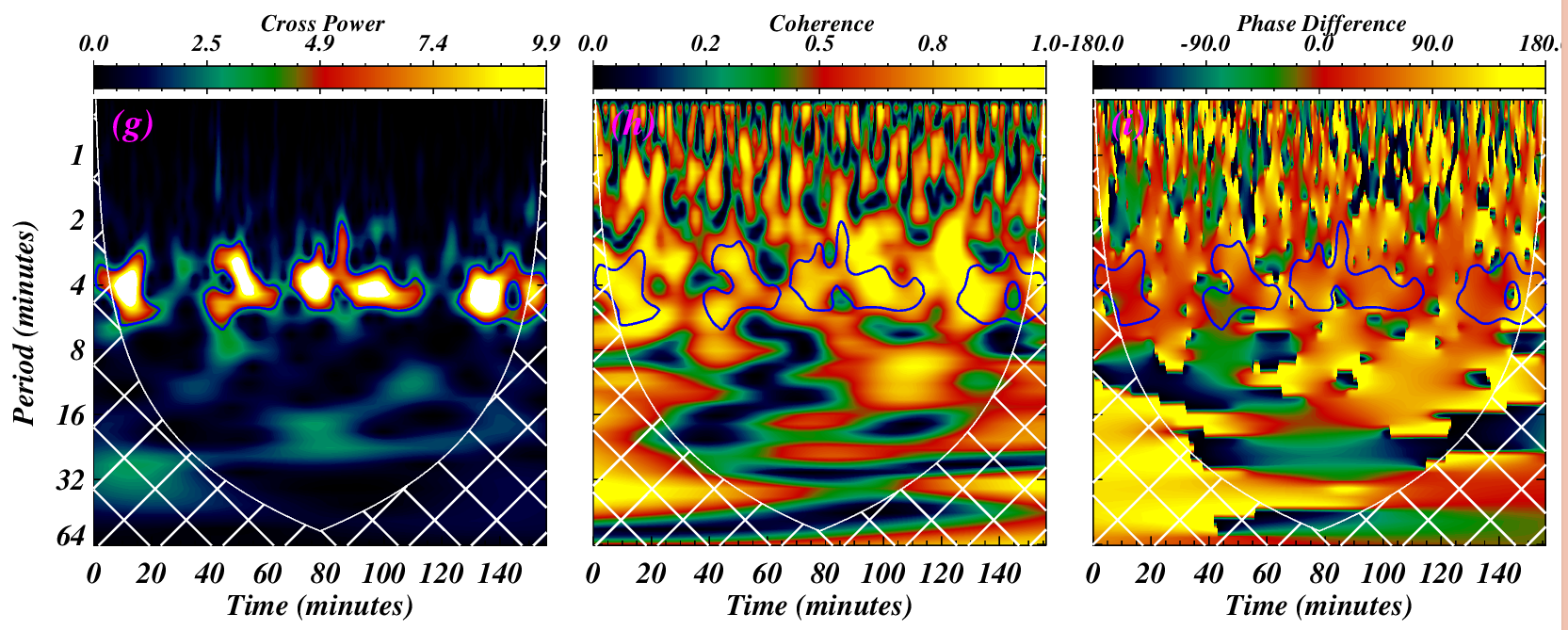}
\caption{The cross power, coherence, and $\Delta$$\phi$, which are estimated using the DTS deduced from Fe~{\sc i} and Mn~{\sc i} for plage location, are displayed in panels (a), (b), and (c). The blue/white contours in all panels are the 95\% confidence levels, and the red cross-hatched area is the COI. The same are displayed for Mn~{\sc i} and Mg~{\sc ii}~k2r (panels (d), (e), and (f)) and Mg~{\sc ii}~k2r and Mg~{\sc ii}~k3 (panels (g), (h), and (i)).}
\label{fig:fig_cross}
\end{figure}

To understand the propagation of waves among these heights in the plage region, we applied detailed cross-wavelet analysis (i.e., estimation of cross power, coherence, and phase difference ($\Delta$$\phi$)) between the pairs of DTS originating from two different heights. From a particular location, the cross-wavelet analysis for different height combinations is shown in Figure~\ref{fig:fig_cross}, namely,  (1) Fe~{\sc i} and Mn~{\sc i} (top row), Mn~{\sc i} and Mg~{\sc ii} k2r (middle row), and Mg~{\sc ii} k2r and Mg~{\sc ii} k3 (bottom row). Panel (a) shows the cross-wavelet power between DTS of Fe~{\sc i} and Mn~{\sc i}, and the bright patch shows the significant common power as it is enclosed by the 95\% confidence levels (see blue contours). The cross power is mainly concentrated between the periods $\approx$ 4 and $\approx$ 8 min. The coherence map for the same is shown in panel (b), and the blue contour is the 95\% confidence level (i.e., same as blue contours in panel (a)). It can be seen that coherence is significant (i.e., greater than 0.8) within the significant cross-power regions. Furthermore, the phase difference is shown in panel (c), and the blue contour corresponds to the 95\% confidence level. Within the region of significant cross-power and coherence, most of the $\Delta$$\phi$ values are positive. In all three panels, the white cross-hatched area is the COI. 
Similarly, the cross power, coherence, and $\Delta$$\phi$ maps for Mn~{\sc i} and Mg~{\sc ii} k2r are displayed in panels (d), (e), and (f), respectively. The same is displayed in panels (g), (h), and (i) for Mg~{\sc ii} k2r and Mg~{\sc ii} k3. For both line pairs (i.e., Mn~{\sc i}{--}Mg~{\sc ii} k2r and Mg~{\sc ii} k2r{--}Mg~{\sc ii} k3), the common cross power, significant coherence, and phase lie within the period range of 3 to 6 minutes.\\

\begin{figure}
\centering
\includegraphics[trim=1.0cm 0.3cm 0.0cm 0.9cm, scale=0.9]{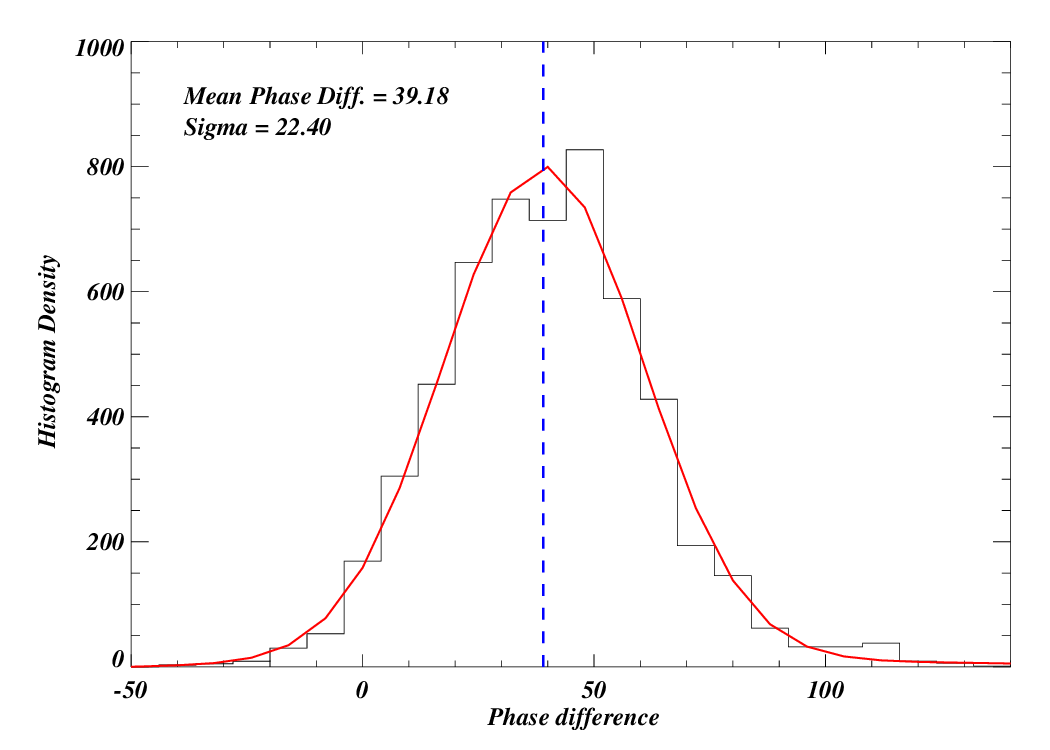}
\caption{The figure shows the histogram of averaged $\Delta$$\phi$ (i.e., phase difference) from all locations (between the heights Fe~{\sc i} and Mn~{\sc i}) at the period of 3.39 min. Further, the Gaussian is fitted over this histogram (see red curve), and the mean $\Delta$$\phi$ and standard deviation (i.e., 1-sigma) are estimated, as mentioned in the figure. The vertical blue line lies at the mean $\Delta$$\phi$ value.}
\label{fig:phase_hist}
\end{figure}

\begin{figure}
\centering 
\includegraphics[trim=1.0cm 0.5cm 2.0cm 0.8cm, scale=0.90]{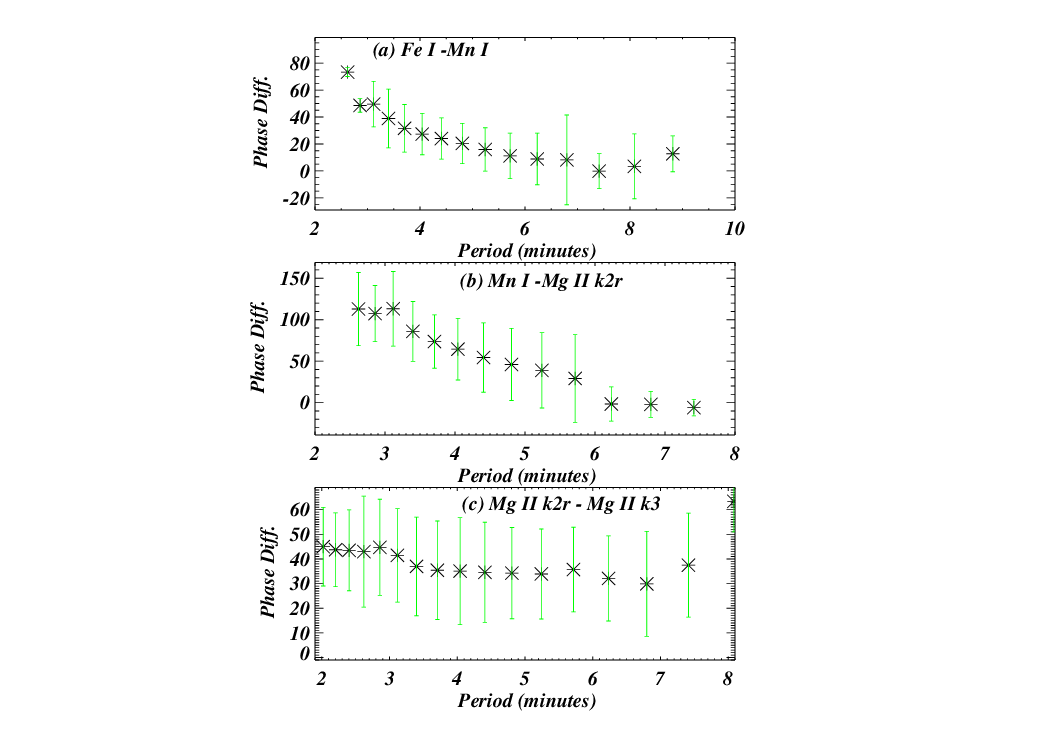}
\caption{The mean $\Delta$$\phi$ is plotted against the period of the waves at different heights in solar plage. Panel (a) shows the variations of the $\Delta$$\phi$ (between the formation heights Fe~{\sc i}  and Mn~{\sc i}) with period, and the $\Delta$$\phi$ is decreasing with period. The same are displayed for Mn~{\sc i} and Mg~{\sc ii}~k2r and Mg~{\sc ii}~k2r and Mg~{\sc ii}~k3 panels (b) and (c), respectively.}
\label{fig:cutoff}
\end{figure}

The $\Delta$$\phi$ deduced between the two heights can tell us whether the waves are moving upward or downward. If $\Delta$$\phi$ is zero, then we consider the wave to be a standing/evanescent wave. To consider only reliable values of $\Delta$$\phi$ for our analysis, we have considered three different criteria, and they are as follows
\begin{itemize}
    \item At first, the cross power should be outside the COI, as the significant power within the COI is ambiguous.  
    \item Next, the cross power should be within the confidence level of 95\%, i.e., the region should be within the blue contour (see panel (a), (d), and (g) of Figure~\ref{fig:fig_cross}).  
    \item The zero value of coherence means the two DTS are completely incoherent, while the coherence value of one reflects complete coherence. In previous works, the coherence value of 0.6 is considered as a reliable threshold value for the coherence between two time-series (e.g., \citealt{2018MNRAS.479.5512K, 2020A&A...634A..63K, 2024ApJ...966..187S}). While some other works report that the coherence above 0.7 can be considered as a reliable threshold value (e.g., \citealt{2017ApJS..229...10J, 2022MNRAS.517..458S}). Therefore, in the present work, we have chosen only those locations where the coherence value is greater than or equal to 0.7.   
\end{itemize}
By applying the conditions described above, for each period, the reliable $\Delta$$\phi$ is collected at each time step from all the plage locations. We have a sufficient number of $\Delta$$\phi$ values in a certain range of periods, for instance, the total number of phase values is 5507 at a period of 3.39 minutes. Further, the histogram of $\Delta$$\phi$ at a period of 3.39 minutes is shown in Figure~\ref{fig:phase_hist}. {\bf Here, it is important to note that we have estimated the cross-power between two DTS, and then, with the significance level of 95\%, a statistical test is performed to find out the significant cross-power in the time-period cross-power map. As statistical tests can produce false positives (Type I errors), we discuss their potential impact on our results. In this case, a particular location has 37260 points (i.e., 570 (time)$\times$60 (period) = 37620), therefore, a maximum of 5\% (i.e., 0.05$\times$37620 = 1881) of total points can be false positives. In the final analysis, we have not included all the points, because we used points inside the 95\% significant power. For instance, from a particular location, we found that 3669 points have significant cross power, so only a maximum of 183 points (i.e., 0.05$\times$3669) might exist by chance (i.e., are unreal). Further, the same significant locations (i.e., 3669) have been used to gather the significant coherence and phase difference values, i.e., we have not employed the statistical tests again. So, we can say the same number of unreal points exist in coherence and phase difference values. Here, again, we mention that we have used histograms to estimate the mean values of phase difference, and these few unreal values (i.e., 183 out of 3669) are very unlikely to affect our results.} Next, the single Gaussian is fitted over this histogram (in red), and the mean $\Delta$$\phi$ and Gaussian width are obtained from this fit. These values (i.e., mean $\Delta$$\phi$ and Gaussian width) are mentioned in the same Figure. The similar exercise is performed for all the periods, and the mean $\Delta$$\phi$ and Gaussian width are estimated.
Finally, Figure~\ref{fig:cutoff} presents the mean phase difference ($\Delta$$\phi$) as a function of period for the different height combinations considered in this study, namely Fe~{\sc i}{--}Mn~{\sc i} (panel a), Mn~{\sc i}{--}Mg~{\sc ii}~k2r (panel b), and Mg~{\sc ii}~k2r{--}Mg~{\sc ii}~k3 (panel c). The green error bars correspond to 1-$\sigma$ at the respective period. To implement height stratification in our analysis, we have used these height combinations, and note that Mg~{\sc ii}~k2r{--}Mg~{\sc ii}~k2v is not considered as the height difference is considerably very small.\\

In Figure~\ref{fig:cutoff}(a), the mean $\Delta$$\phi$ decreases with increasing period. Furthermore, the mean $\Delta$$\phi$ is positive below $\sim$ 6 minutes, i.e., all periods shorter than 6 minutes are propagating upward from the formation height of Fe~{\sc i} to the formation height of Mn~{\sc i}. After 6 minutes, the mean phase difference is almost equal to zero, implying the presence of stationary/evanescent waves.
Similar behavior of $\Delta$$\phi$ exists between the heights Mn~{\sc i} and Mg~{\sc ii}~k2r (Figure~\ref{fig:cutoff}(b)). However, the initial $\Delta$$\phi$ is higher (i.e., 120\textdegree) than the initial $\Delta$$\phi$ between Fe~{\sc i} and Mn~{\sc i} ($\sim$70\textdegree in panel (a)). After $\sim$6 minutes, the $\Delta$$\phi$ becomes zero, similar to the previous case (Figure~\ref{fig:cutoff}(a)). Before 6 minutes, all the $\Delta$$\phi$ are positive, and it means that such waves are propagating upward. Lastly, we mention that $\Delta$$\phi$ between Mg~{\sc ii} k2r and Mg~{\sc ii} k3 also decreases with period, like in the previous height combinations (Figure~\ref{fig:cutoff}(a) and Figure~\ref{fig:cutoff}(b)). But, the $\Delta$$\phi$ is the least for this height combination. Further, the rate of decrease of $\Delta$$\phi$ with period is also lower than the previous ones, i.e., the $\Delta$$\phi$ decreases from 45\textdegree to around 30\textdegree in the period range of 2.0 to 7.0 minutes. All the $\Delta$$\phi$ have positive values, i.e., all the waves are propagating upward from 2.0 to more than 7.0 minutes. 
\subsection{Propagation Speed of Wave at Different Heights}
We can estimate the wave travel time using the $\Delta$$\phi$ (in radian) and particular frequency, as explained by \cite{2017ApJS..229...10J} and \cite{2020A&A...634A..63K}, see the equation below.
\begin{equation}
    \tau = {\Delta \phi \over 2 \pi f} 
    \label{eq:eq1}
\end{equation}
Where $\tau$, $\Delta$$\phi$, and f are wave travel time, phase difference, and frequency, respectively. Further, the wave travel time and height difference can be used to estimate the propagation speed of waves. Please note that the values of $\Delta$ $\phi$ at 3.11 min (i.e., 5.36 mHz) are $\sim$48.55\textdegree $\pm 5.04\textdegree $, $\sim$113.21\textdegree $\pm 44.95\textdegree$, $\sim$41.47 \textdegree $\pm 18.98\textdegree$ within the photosphere (i.e., Fe~{\sc i}{--}Mn~{\sc i}), photosphere-chromosphere (i.e., Mn~{\sc i}{--}Mg~{\sc ii} k2r), and chromosphere (i.e., Mg~{\sc ii} k2r{--}Mg~{\sc ii} k3), respectively. Hence, the wave travel times, according to equation~\ref{eq:eq1}, are 25.21$\pm 2.62$, 58.63$\pm 23.36$, and 21.54$\pm 9.86$ s, respectively. Now, for these height combinations, we know that height differences are 0.33 Mm (between Fe~{\sc i} and Mn~{\sc i}), 0.57 Mm (between Mn~{\sc i} and Mg~{\sc ii} k2r), and 0.50 Mm (between Mg~{\sc ii} k2r and Mg~{\sc ii} k3). Therefore, the propagation speeds are 13.24$\pm$1.38 km/s (range: 11.86 to 14.62 km/s) in the photosphere, 9.72$\pm$2.76 km/s (range: 6.96 to 12.48 km/s) in the photosphere-chromosphere, and 23.21$\pm$7.28 km/s (range: 15.93 to 30.49 km/s) in the chromosphere. Further, we have considered the uncertainty in the formation heights of these lines, as mentioned in the Table~\ref{tab:tab1}, to understand the variations in propagation velocities. We estimate the minimum and maximum height differences for each line pair, and then estimate the propagation speeds for these height differences using the mean travel time (mentioned above). The propagation speeds vary from 7.53 km/s (minimum height difference) to 17.05 km/s (maximum height difference) for Fe~{\sc i} and Mn~{\sc i} (at the photosphere), while the same vary from 2.89 km/s to 16.54 km/s for Mn~{\sc i} and Mg~{\sc ii}k2r (photosphere-chromosphere). For chromosphere (i.e., Mg~{\sc ii}k2r and Mg~{\sc ii} k3), the propagation speeds vary from 4.64 km/s to 41.78 km/s. 

\subsection{Formation Heights of Spectral Lines} \label{sect:fmh} 
The work focuses on the dominant periods at multiple heights in the pore and plage of the solar atmosphere using a few spectral lines. The formation heights of these spectral lines are tabulated in table~\ref{tab:tab1}, and these values of the formation height are appropriate for the network of QS (e.g., \citealt{1981ApJS...45..635V}). However, the formation height of any line depends on the local thermal and magnetic field conditions. We also know that thermal and magnetic field conditions are very different in the pore/plage than in the QS. Therefore, we expect significant differences in the formation heights of these lines among the pore, plage, and QS, and variations in the formation heights of spectral lines across these regions should be discussed.\\

The relative height difference between two spectral lines can be estimated by phase difference (\citealt{2022ApJ...933..109Z, 2024ApJ...972...39K}). The plage, pores, and QS exist in the observed region, as explained in Section~\ref{sect:roi}. In case of pore, the $\Delta$$\phi$ is estimated from each DTS for three different lines (i.e., Mg~{\sc ii} k3, Mg~{\sc ii} k2r, and Mn~{\sc i}) relative to the lowest height line (i.e., Fe~{\sc i}). This way, we have three line pairs, namely, (1) Fe~{\sc i}{--}Mn~{\sc i}, (2) Fe~{\sc i}{--}Mg~{\sc ii} k2r, and (3) Fe~{\sc i}{--}Mg~{\sc ii} k3. Now, for each line-pair, we have a 3-D array of $\Delta$$\phi$(i.e., {\bf 3-D array = [no. of times points (570), no. of period (66), no. of pore locations (30)]}). Further, we have kept only reliable $\Delta$$\phi$ in the 3-D $\Delta$$\phi$ array by applying the three criteria mentioned in the previous section. Now, we have a 3-D array of reliable $\Delta$$\phi$ with missing values at unreliable locations. At each period, we extracted all the reliable $\Delta$$\phi$, and produced the histogram of the reliable $\Delta$$\phi$. The histogram is fitted with the Gaussian function to estimate the mean $\Delta$$\phi$ and 1-$\sigma$ error in the mean $\Delta$$\phi$. The same exercise is applied to other line pairs of pores, and then again, the whole procedure for plage and QS is repeated.

\begin{figure}
\centering
\includegraphics[trim=0.0cm 0.0cm 0.0cm 0.5cm,scale=0.98]{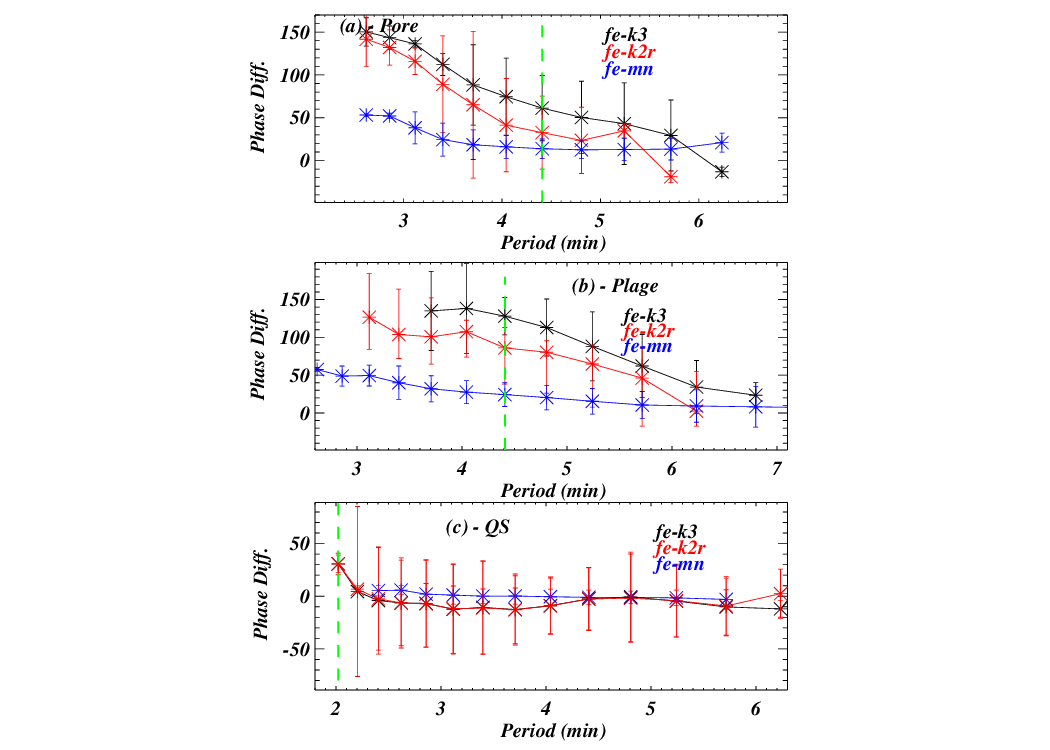}
\caption{The mean $\Delta$$\phi$ at different height differences is plotted against the period for pore (panel a), plage (panel b), and QS (panel c). The black curve at each panel represents the phase difference between the heights of Fe~{\sc i} and Mg~{\sc ii}~k3. The same profiles between Fe~{\sc i} and Mg~{\sc ii}~k2r are shown in red, and between Fe~{\sc i} and Mn~{\sc i} are shown in blue. The green dashed line represents the period at which the propagation speeds are estimated. }
\label{fig:fmh}
\end{figure}
Panel (a) of Figure~\ref{fig:fmh} shows the mean $\Delta$$\phi$ with period for Fe{\sc i}{--}Mg~{\sc ii}~k3 (black profile), Fe{\sc i}{--}Mg~{\sc ii}~k2r (red profile), and Fe{\sc i}{--}Mn~{\sc i} (blue profile). For all three line pairs, the $\Delta$$\phi$ decreases with wave period: an obvious fact of waves in the solar atmosphere (e.g., \citealt{2020A&A...634A..63K, 2022ApJ...933..109Z, 2024ApJ...972...39K}). On average, $\Delta$$\phi$ is maximum for Fe{\sc i}{--}Mg~{\sc ii}~k3 (black profile) because the height difference is maximum (i.e., 1.90 Mm-0.50 Mm = 1.40 Mm). While $\Delta$$\phi$ is minimum for Fe{\sc i}{--}Mn~{\sc i} (blue profile) because the height difference is minimum(i.e., 0.83 Mm-0.50 Mm = 0.33 Mm). And, the $\Delta$$\phi$ for Fe{\sc i}{--}Mg~{\sc ii}~k2r (red profile) lies in between maximum and minimum height difference, as the height difference (i.e., 1.40 Mm-0.50 Mm = 0.90 Mm)is greater than the minimum height difference (i.e., 0.33 Mm)and smaller than the maximum difference (i.e., 1.40 Mm). The same behavior of $\Delta$$\phi$ with period is observed in plage (panel (b), Figure~\ref{fig:fmh}). In contrast, the QS has different behavior than the plage and pore (panel (c), Figure~\ref{fig:fmh}). The first difference is that the $\Delta$$\phi$ for minimum height difference (i.e., between Fe~{\sc i} and Mn~{\sc i}) is nearly zero (see blue profile in panel (c)). The second difference is that the $\Delta$$\phi$ is the same for Fe~{\sc i}{--}Mg~{\sc ii}~k2r (red profile) and Fe~{\sc i}{--}Mg~{\sc ii}~k3 (black profile) in the QS. The third difference is that $\Delta$$\phi$ becomes negative after 2.4 minutes and continuously decreases up to 4 minutes, and later, $\Delta$$\phi$ increases.

As explained in the previous section, we can estimate the travel time of any wave period using $\Delta$$\phi$ (e.g., \citealt{2020A&A...634A..63K, 2022ApJ...933..109Z}), see Equation~\ref{eq:eq1}. Next, we can also estimate the wave (phase) velocity (or propagation speed) at that period if we know the difference in the formation heights of both lines of the line pair, i.e., v$_{ph}$ = $\Delta$ H/$\tau$. Here, v$_{ph}$ and $\Delta$H are the wave velocity and formation height difference for any particular line-pair. The formation heights mentioned in table~\ref{tab:tab1} are suitable for the QS (e.g., \citealt{1981ApJS...45..635V}). Therefore, these formation heights might not be suitable for pore/plage, as the physical conditions are very different in these regions than in QS. Therefore, we try to understand the relative height difference in the pore and plage.
\begin{table*}
\centering
\caption{Travel time and phase velocity for different line pairs for plage.}
\begin{tabular}{|c|c|c|c|c|c|c|}  

\hline
Sr.No & Line-Combination & $\Delta$$\phi$ (rad) & 2$\pi$ $\nu$ & QS Height Difference (Mm) & $\tau$ (s) & Propagation Speed (km/s)\\
  \hline

1 &  Fe~{\sc i}-Mn~{\sc i} & 0.423  & 0.02374 & 0.330 (0.83-0.50) & 17.81  & 18.52 \\
\hline
2 & Fe~{\sc i}-Mg~{\sc ii} k2r & 1.501  & 0.02374 & 0.900 (1.4-0.50) & 63.23  & 14.23 \\
\hline
3 & Fe~{\sc i}-Mg~{\sc ii} k3 & 2.234  & 0.02374 & 1.40 (1.9-0.50) & 94.10  & 14.87 \\

\hline
\end{tabular}
\label{tab:tab_plage}
\end{table*}
In case of plage, we have estimated $\tau$ and propagation speed (v$_{ph}$) at a particular period of 264.459 s. The vertical green dashed line is drawn at this period in panel (b) of Figure~\ref{fig:fmh}. All estimations related to propagation speed calculations are tabulated in the Table~\ref{tab:tab_plage}. The propagation speed in the chromosphere is around 14 km/s (i.e., 14.23 km/s for Mg~{\sc ii} k2r and 14.87 km/s for Mg~{\sc ii} k3), which is lower. Please note these propagation speeds in plages are estimated using the height difference, which is suitable for QS; therefore, the lower velocity suggests that the height difference is underestimated in the plage. The propagation speed in the chromosphere above plage should vary from 15 to 20 km/s (e.g., \citealt{2020A&A...634A..63K}). Using the estimated travel time ($\tau$), the lower (15 km/s) and upper velocities (20 km/s) can be achieved, if the formation heights of Mg~{\sc ii} k3 are 1411.50 km (i.e., 94.10$\times$15.0) and 1882.0 km (i.e., 94.10$\times$20) relative to the formation height of Fe~{\sc i} (i.e., 500 km). \cite{2020A&A...634A..63K} have shown excess velocity around 30 km/s in the plage, and as per this velocity Mg~{\sc ii} k3 is forming at a height of 2728.9 km (i.e., 94.10$\times$30) relative to the formation height of Fe~{\sc i}. Similarly, the same range of velocities can be achieved for Mg~{\sc ii} k2r if the formation heights vary from 948.45 km to 1264.6 km/s relative to the formation height of Fe~{\sc i}. On the contrary, the propagation speed at the photosphere in plage is higher than in the chromosphere, which is unusual. It means the formation height difference between Fe~{\sc i} and Mn~{\sc i} is overestimated. Even if we consider the lower range of propagation speed from \cite{2020A&A...634A..63K}, the Mn~{\sc i} should form at a height of 267.15 km relative to Fe~{\sc i}.\\
\begin{table*}
\centering
\caption{ Travel time and phase velocity for different line pairs for pore.}
\begin{tabular}{|c|c|c|c|c|c|c|}  

\hline
Sr.No & Line-Combination & $\Delta$$\phi$ (rad) & 2$\pi$ $\nu$ & QS Height Difference (Mm) & $\tau$ (s) & Propagation Speed (km/s)\\
  \hline
1 &  Fe~{\sc i}-Mn~{\sc i} & 0.2401  & 0.02374 & 0.330 (0.83-0.50) & 10.12  & 32.60 \\
\hline
2 & Fe~{\sc i}-Mg~{\sc ii} k2r & 0.569  & 0.02374 & 0.900 (1.4-0.50) & 23.96  & 37.56 \\
\hline
3 & Fe~{\sc i}-Mg~{\sc ii} k3 & 1.068  & 0.02374 & 1.40 (1.9-0.50) & 44.98  & 31.12 \\
\hline
\end{tabular}
\label{tab:tab_pore}
\end{table*}
Similar to plage, we estimated $\tau$ and wave velocity (v$_{ph}$) for pore at the same period of 264.459 s (see table~\ref{tab:tab_pore}), and the green dashed line in panel (a) of Figure~\ref{fig:fmh} shows the location of this period. All the related calculations are tabulated in Table~\ref{tab:tab_pore}. The pore has more than double propagation speeds at the photosphere and chromosphere than solar plage, i.e., compare the propagation speeds in the last columns of the table~\ref{tab:tab_plage} and~\ref{tab:tab_pore}. But generally, we know the pore should have lower propagation velocities than solar plage, because the pore has a lower temperature than solar plage. Hence, most probably, the height difference is overestimated, leading to the excess propagation velocities. Again, for estimating propagation speeds in the pore, the height difference is taken from the QS (\cite{1981ApJS...45..635V}). Even if we assume that the propagation speeds in the pore are as low as those in solar plages, corresponding to these propagation velocities, the formation heights of Mn~{\sc i}, Mg~{\sc ii} k2r, and Mg~{\sc ii} k3 are 187.42 km (18.52$\times$10.12), 340.95 km (14.23$\times$23.96), and 668.85 km (14.87$\times$44.98) relative to the formation height of Fe~{\sc i}. In reality, the propagation speeds would be even less than the values used here in the estimation (i.e., plage propagation speeds). Note that \cite{2015ApJ...806..132G} have reported that propagation velocity in the photosphere above a pore is around 5 km/s, which results in the Mn~{\sc i} line forming at a height of around 50.60 km relative to the Fe~{\sc i}. However, \cite{2015ApJ...806..132G} have also reported a high propagation speed of 15 km/s in the photosphere above the pore, and as per this speed, the Mn~{\sc i} line should form at a height of 151.8 km from the formation height of Fe~{\sc i}\\

Further, we performed a similar analysis for the QS (panel (c), Figure~\ref{fig:fmh}). The $\Delta$$\phi$ is close to zero (i.e., varying from -1.0 to +5.0) for Fe~{\sc i}{--}Mn{\sc i} line pair (blue profile, Figure~\ref{fig:fmh}). We cannot expect standing waves within the photosphere, as can be expected between photosphere and chromosphere (i.e., Fe{\sc i}{--}Mg~{\sc ii}~k2r and Fe{\sc i}{--}Mg~{\sc ii}~k3). Hence, most probably, the zero $\Delta$$\phi$ at all the periods in QS implies that Fe~{\sc i} and Mn~{\sc i} are forming at the same height. Next, $\Delta$$\phi$ of Mg~{\sc ii}~k2r and Mg~{\sc ii}~k3 relative to Fe~{\sc i} are also the same at almost all periods (red and black profiles in panel (c), Figure~\ref{fig:fmh}). It suggests that most probably Mg~{\sc ii k2r} and Mg~{\sc ii} k3 are forming at the same height relative to Fe~{\sc i}. The $\Delta$$\phi$ is around 0.5365 rad at a period of 121.255 s for both line pairs, then the travel time ($\tau$) of this period to reach the formation heights of Mg~{\sc ii} k2r is around 10.36 s. Hence, the propagation speed is 86.87 km/s (i.e., 900 km/10.36 s). It is extremely high propagation speed in such a low magnetic field and temperature region. Here, it is important to note QS, which we have used in this study, does not contain bright emission and a strong magnetic field, as networks usually have. Therefore, it is an internetwork region of the QS, i.e., a magnetic field-free region (panel (e), Figures~\ref{fig:ref_fig}). Hence, we mention that in this internetwork QS Fe~{\sc i} and Mn~{\sc i} are forming nearly at the same height, and similarly, the Mg~{\sc ii}~k2r and Mg~{\sc ii}~k3 are also forming at the same height.
\section{Summary $\&$ Discussions}
The propagation of waves among different layers of the solar atmosphere is an important topic of solar physics research, and it (wave propagation) has been investigated for different physical conditions, i.e., for different regions within the solar atmosphere, including solar plage (e.g., \citealt{2003ApJ...595L..63D, 2006ApJ...640.1153C, 2009ApJ...692.1211C, 2015LRSP...12....6K, 2017ApJS..229...10J, 2018MNRAS.479.5512K, 2020A&A...634A..63K, 2022MNRAS.517..458S, 2023LRSP...20....1J, 2024ApJ...966..187S}). The present work is dedicated to understanding how the (dominant) wave periods change with height between the photosphere and the chromosphere within the solar plage and pore. Further, we also investigated their (waves) propagation from one height to another in plage regions. To do so, we have tried to stratify different heights between the photosphere and the chromosphere and understand the propagation at these heights. Lastly, we have investigated the variations in the formation heights of different spectral lines in different physical conditions. The main findings of the work are summarized below.
\begin{itemize}
    \item The mean dominant periods within the solar plage are 5.0, 4.40, 4.40, 4.70, and 4.70 at the formation heights of Fe~{\sc i}, Mn~{\sc i}, Mg~{\sc ii} k2r, Mg~{\sc ii} k2v, and Mg~{\sc ii} k3, respectively. It means that the dominant period in plages shows no significant change at different heights between the photosphere and chromosphere (Figure~\ref{fig:domin_correl}(a)).
    
    \item While, in the case of pores, the mean dominant periods are 4.20, 4.0, 3.80, 3.0, and 3.2 minutes at the formation heights of Fe~{\sc i}, Mn~{\sc i}, Mg~{\sc ii} k2r, Mg~{\sc ii} k2v, and Mg~{\sc ii} k3, respectively. Hence, pores show (almost) a linear decrease in dominant periods from the photosphere to chromosphere with a strong Pearson coefficient of -0.916 (Figure~\ref{fig:domin_correl}(b)).

    \item Wave propagation analysis, using cross wavelet analysis, reveals that for all three height combinations (i.e., (1) Fe{\sc i}{--}Mn{\sc i}, (2) Mn{\sc i}{--}Mg{\sc ii} k2r, and (3) Mg{\sc ii} k2r{--}Mg{\sc ii} k3), the $\Delta$$\phi$ decreases with the wave period. On average, the second height combination has the highest $\Delta$$\phi$ (Figure~\ref{fig:cutoff}(b)), while the third height combination has the lowest $\Delta$$\phi$ (Figure~\ref{fig:cutoff}(c)). 
  
    \item The $\Delta$$\phi$ is positive for the periods below $\sim$6 minutes for the Fe~{\sc i}{--}Mn{\sc i} and Mn~{\sc i}{--}Mg~{\sc ii} k2r height combinations, i.e., all these wave periods are propagating upward from the photosphere to middle chromosphere within the solar plages. While after these periods (i.e.,~$\sim$6 minutes), the $\Delta$$\phi$ spectrum become flat at around 0\textdegree~(Figure~\ref{fig:cutoff}(a) and Figure~\ref{fig:cutoff}(b)). On the contrary, all the waves are propagating upward for the third height combination (i.e., Mg~{\sc ii} k2r and Mg~{\sc ii} k3), see Figure~\ref{fig:cutoff}(c). 
    
    \item We estimated the propagation speeds of waves within the solar photosphere, photosphere-chromosphere, and chromosphere for a frequency of 5.36 mHz (i.e., 3.11 minutes). The propagation speeds are 13.24$\pm$1.38, 9.472$\pm$2.76, and 23.21$\pm$7.28 km/s in the photosphere, middle chromosphere, and upper chromosphere, respectively.

    \item Lastly, qualitatively, with the help of $\Delta$$\phi$, we have attempted to understand the variations in formation heights of Mn~{\sc i}, Mg~{\sc ii} k2r, and Mg~{\sc ii} k3 relative to the formation height Fe~{\sc i} in the pore and plage. A significant difference in the formation heights exists between pore and plage.
\end{itemize}    

In various features of the solar atmosphere (e.g., sunspots, pores, QS, and CH), the power in 3 min (i.e., 5.5 mHz) in the photosphere is less than the power in 5 min waves (i.e., 3.3 mHz) at the photosphere. While 3-minutes has more power than 5 minutes in the solar chromosphere of these solar atmospheric features (e.g., \citealt{1982ApJ...253..367L, 2009ApJ...692.1211C, 2015LRSP...12....6K, 2018MNRAS.479.5512K, 2022MNRAS.517..458S,2024ApJ...966..187S,2025ApJ...987..161K}). However, the maximum power lies at a period of 5 minutes within the solar photosphere and chromosphere above the solar plages/networks (e.g., \citealt{2003ApJ...595L..63D, 2009ApJ...692.1211C, 2020A&A...634A..63K}). We have also found that the dominant periods are around 5 minutes at all atmospheric heights from the photosphere to the chromosphere in the solar plage (see Table~\ref{tab:tab2}). It means that the dominant period is not changing (i.e., increasing or decreasing) with height in the solar plages. However, recently, in the QS, \cite{2025ApJ...987..161K} investigated the dominant period with height, and it is reported that the dominant period decreases from 271 s (at 0.5 Mm) to 167 s (at 2.2 Mm) with height. It means the behaviour of the dominant period in plages is different from other features.\\

However, unlike plage regions, we observe a linear decrease in the dominant period from 5 (at the photosphere) to 3 min (in the chromosphere) in the pore regions. The pores are actually the umbra without penumbra, and it is very well established that the photosphere (chromosphere) of the umbra is dominated by 5-(3-minutes) (e.g., \citealt{2009ApJ...692.1211C, 2014ApJ...786..137T, 2021ApJ...906..121K}). Therefore, the dominant period should decrease systematically with height from the photosphere to the chromosphere, which is reported in the present work.\\

It is reported that the $\Delta$$\phi$ decreases with wave periods (e.g., \citealt{2017ApJS..229...10J, 2020A&A...634A..63K}), and the same is reported in the present work for different height combinations in the solar plages (see Figure~\ref{fig:cutoff}). Further, using $\Delta$$\phi$ and height difference, the wave propagation speeds are estimated. The propagation speed depends on the type of wave and in-situ physical conditions (i.e., temperature, density, and magnetic field). \cite{2017ApJS..229...10J} have reported the speed of 31$\pm$2 km/s of longitudinal wave over a magnetic bright point in the chromosphere. Further, \cite{2020A&A...634A..63K} reported that the speed is 29 km/s above the plages in the chromosphere/TR, and they reported that the estimated speed matches the sound speed of TR. Therefore, the waves were interpreted as slow magnetoacoustic waves. In the present work, it is found that the propagation speed of waves in the photosphere (i.e., 13.24$\pm$1.38 km/s) is higher than the typical sound speeds of the photosphere (i.e., 6-9 km/s). However, if the uncertainty in height is taken into account, then the propagation speed varies from 7.53 to 17.05 km/s. On the other hand, the propagation speed at the middle chromosphere (i.e., 9.72$\pm$2.76 km/s) is consistent with the sound speed of the middle chromosphere (i.e., 10-15 km/s). Again, the propagation speed of waves in the upper chromosphere (i.e., 23.21$\pm$7.28 km/s) is slightly higher than the sound speed of the upper chromosphere (i.e., 15-20 km/s). Next, we mention that plages have higher temperatures than the QS (\citealt{2015ApJ...809L..30C}); therefore, we can expect higher sound speeds in the atmosphere above plages. Most importantly, the height difference considered in the propagation speed estimation of the wave is not very accurate. Specifically, in the strong magnetic field regions like solar plages, the formation height can change significantly (as discussed in the last paragraph), and as a result, the propagation speeds change accordingly. Hence, keeping these aspects in mind, we can say that the waves are slow magnetoacoustic waves, as the propagation speeds are close to the sound speed if we consider the above-described facts (i.e., higher temperature and variations in the formation heights).

Another crucial piece of information about these slow MHD waves (i.e., cutoff frequency) can be inferred from the variations of $\Delta$$\phi$ with periods (e.g., \citealt{2006ApJ...647L..77M, 2006PhRvE..73c6612M, 2016ApJ...819L..23W, 2017ApJS..229...10J, 2018MNRAS.479.5512K, 2021ApJ...906..121K, 2024ApJ...966..187S, 2006MNRAS.372..551S}). Observationally, in most of the works, the period corresponding to zero phase difference (in $\Delta$$\phi$ vs period diagram) is considered as the cutoff period (e.g., \citealt{2016ApJ...819L..23W, 2018A&A...617A..39F, 2018MNRAS.479.5512K, 2024ApJ...966..187S, 2006MNRAS.372..551S}). In the present work, we have found that consistently $\Delta$$\phi$ is zero after around 6 minutes at the photosphere (Figure~\ref{fig:cutoff}(a)) and in the middle chromosphere (Figure~\ref{fig:cutoff}(b)). While in the upper chromosphere, the $\Delta$$\phi$ decreases very slowly from 2.0 to 7.0 minutes, i.e., $\Delta$$\phi$ decreases from 44\textdegree to 32\textdegree within this period range (Figure~\ref{fig:cutoff}(c)). In this case, all the waves are propagating from the middle to the upper chromosphere, and like previous cases, we don't see a flat spectrum after a particular period. Hence, 6 minutes can be considered the cutoff period at photospheric and middle chromospheric heights. And, in the present work, we didn't observe the cutoff from the middle to the upper chromosphere in this period range. 
Lastly, we mention $\Delta$$\phi$ is positive before the cutoff period (i.e., 6 minutes); therefore, all the wave periods below 6 minutes, including 5 minutes, reach from the photosphere to the middle chromosphere. And, all the waves till the period of 7.0 minutes propagate from the middle to the upper chromosphere. Previous works have also shown the dominant 5-minute period from the photosphere propagates up to the TR in the solar plages (e.g., \citealt{2003ApJ...595L..63D, 2005ApJ...624L..61D, 2009ApJ...692.1211C, 2020A&A...634A..63K}), and therefore our results are consistent with the previous findings.\\

Lastly, we have worked on the variations of formation heights of these spectral lines, using the $\Delta$$\phi$ profile with period. We estimate the $\Delta$$\phi$ vs. period profile for Mg~{\sc ii} k3, Mg~{\sc ii} k2r, and Mn~{\sc i} relative to Fe~{\sc i} (i.e., the lower formation height). These profiles are estimated for pore, plage, and QS. The propagation speeds come very low in the chromosphere above the plages, which suggests that formation heights of Mg~{\sc ii} k2r and Mg~{\sc ii} k3 are significantly underestimated. While in the plage, the formation height of Mn~{\sc i} (i.e., photospheric lines) is overestimated, as the propagation speed is higher at the photosphere than the propagation speeds in the chromosphere. In pores, the formation heights of Mn~{\sc i}, Mg~{\sc ii} k2r, and Mg~{\sc ii} k3 are overestimated, as the propagation speeds are very high (two times the propagation speeds in plage). It suggests that the formation height in the pore should be decreased at least by half to get the propagation speeds comparable to the propagation speeds of the plage. The results are even more interesting regarding the formation heights in the QS, i.e., most probably, Fe~{\sc i} and Mn~{\sc i} are forming nearly at the same height, as the $\Delta$$\phi$ is nearly zero in all the periods. Also, Mg~{\sc ii}k2r and Mg~{\sc ii} k3 are forming at the same height in the chromosphere, as they have almost equal $\Delta$$\phi$ profile relative to Fe~{\sc i} line.
\section*{Acknowledgments}
We acknowledge the constructive comments/feedback on the manuscript by the anonymous referee, which has improved the quality of the work. IRIS is a NASA small explorer mission developed and operated by LMSAL, with mission operations executed at NASA Ames Research Center and major contributions to downlink communications funded by ESA and the Norwegian Space Centre. We also acknowledge the use of magnetic field observations provided by SDO/HMI.  


\appendix 
\label{append1}
{\section{Big Patches of Bad Spectra in the Observations}
In this observation, IRIS had captured the spectra along the slit for more than 5 hours, i.e., the complete observation has 1128 time steps. 
\begin{figure}
\centering 
\includegraphics[trim=2.0cm 1.8cm 2.0cm 1.0cm, scale=0.92]{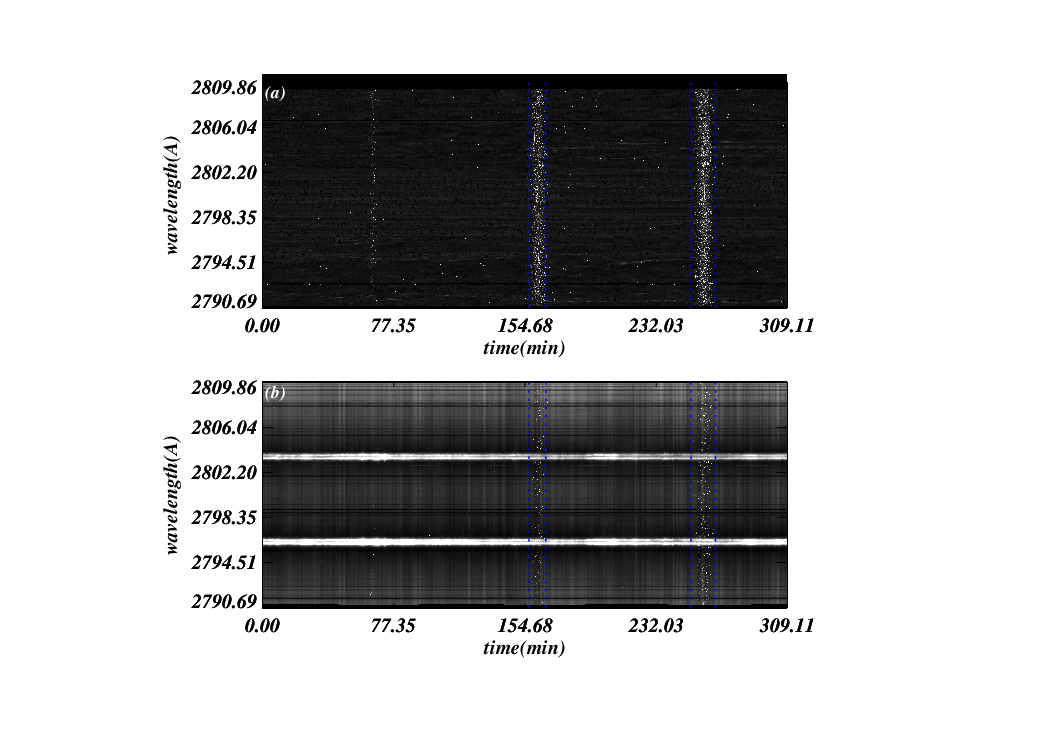}
\caption{Figure shows the wavelength ($\lambda$)-time (t) diagram of the complete observation in the photospheric (panel) and chromosphere window (panel b). The vertical blue dashed lines enclose two big patches, whose spectra are badly affected at all wavelengths. }
\label{fig:append1}
\end{figure}
\begin{figure}
\centering 
\includegraphics[trim=2.0cm 2.0cm 2.0cm 1.5cm, scale=0.92]{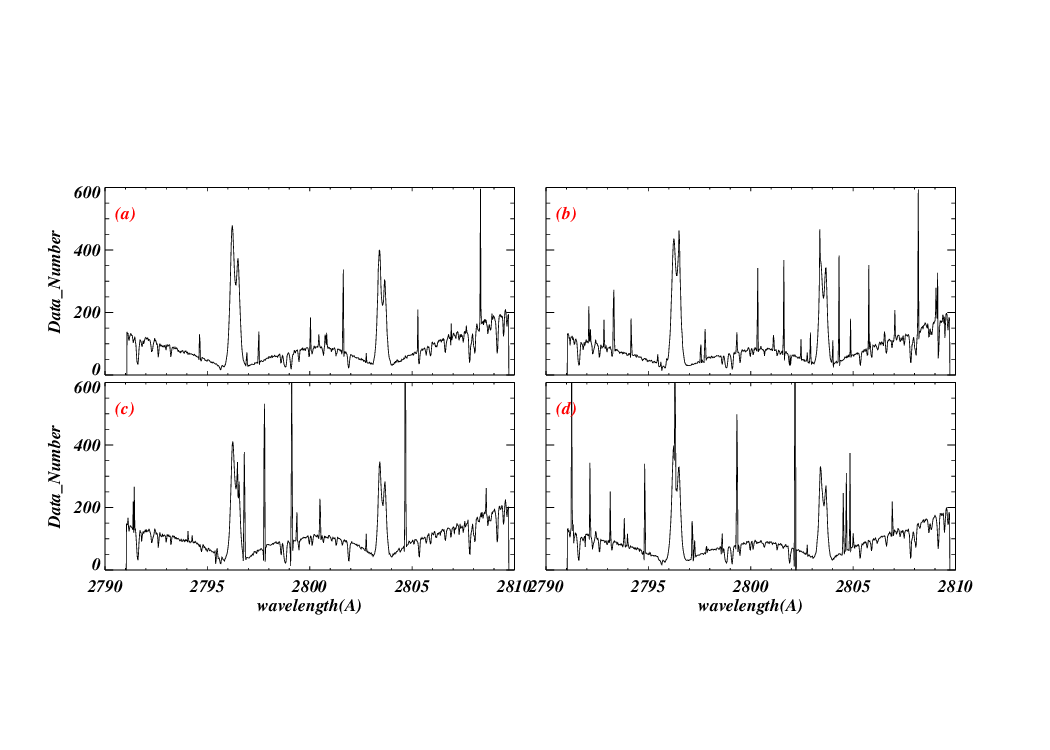}
\caption{This figure shows four spectra from the chromospheric spectral window, and several spikes are clearly visible in these spectra, which affect the Doppler velocity estimation.}
\label{fig:append2}
\end{figure}
The panel (a) of Figure~\ref{fig:append1} shows the t-$\lambda$ image of the photosphere window from one particular Y-location (y = 30.55$"$). The quality of the spectra is good till time 152.53 minutes (i.e., time position = 569), but after this, the white dots (i.e., strong intensity) are distributed all over the complete wavelength range. The existence of white dots lasts till 162.10 minutes, i.e., the bad quality spectra exist for more than 10 minutes (see the region enclosed by blue dashed lines). One more similar patch exists from 245 to 259 minutes (see the region enclosed by blue dashed lines). Similar to panel (a), another t-$\lambda$ image from IRIS Mg~{\sc ii} k window from the same Y-location is displayed in panel (b). The two bright horizontal patches are actually Mg~{\sc ii} k 2796.35~{\AA} and Mg~{\sc ii} k 2803.52~{\AA}. In addition to these bright patches, several dark horizontal lines are also visible, and the photospheric absorption lines exist in this window. Please note all the lines used in this study exist in this spectral window, i.e., from 2790.69~{\AA} to 2809.86~{\AA}. The same patches of bad spectra are also outlined by blue dashed lines.

We have displayed spectra from the NUV window from some time steps in panels (a), (b), (c), and (d) of Figure~\ref{fig:append2}. The multiple sharp spikes are visible in the spectra, and most probably, these spikes are due to the cosmic ray hits, not due to the real phenomena happening within the solar atmosphere. Hence, the spectra are badly affected by the cosmic hits for the time intervals of 10 minutes (i.e., from 152 to 162 minutes) and around 14 minutes (i.e., from 245 to 259 minutes). Therefore, the estimation of the Doppler velocity of various emission/absorption lines from such low-quality spectra would not be reliable. Also, the time gaps of 10 and 14 minutes are huge, and to get the Doppler velocity during these low-quality spectra using interpolation techniques will not be reliable. Therefore, we have considered the Doppler velocity time-series before the occurrence of the first patch of bad spectra, i.e., before 152 minutes.}

\begin{figure}
\centering 
\includegraphics[trim=2.0cm 3.0cm 2.0cm 2.0cm, scale=1.0]{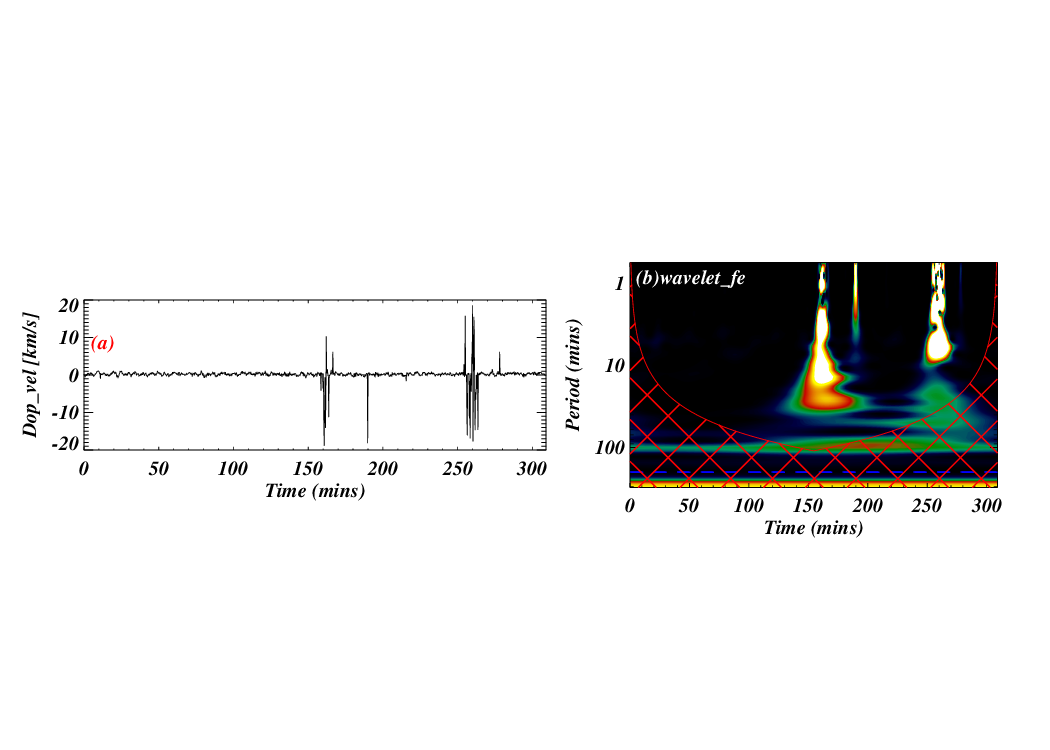}
\caption{The wavelet power map from a particular plage location for the complete time of the observation.}
\label{fig:append3}
\end{figure}

{Lastly, we showed the wavelet power map (Figure~\ref{fig:append3}(b))for a particular DTS (Figure~\ref{fig:append3}(b)) for complete time, including the bad spectra. It is clearly visible that Doppler velocities are wrongly estimated at the time of bad spectra, and due to the maximum power existing at these times only, see the bright regions in panel (b).}
\end{document}